\documentclass[aps,pra,twocolumn,superscriptaddress,amsmath,amssymb]{revtex4-2}
\usepackage{newtxtext,newtxmath}
\usepackage[T1]{fontenc}
\usepackage{graphicx}
\usepackage{bm}
\usepackage{siunitx}
\usepackage{tabularx}
\usepackage[colorlinks=true,citecolor=blue,breaklinks=true,linkcolor=blue,
linktocpage=true,urlcolor=blue,pagebackref=false]{hyperref}
\usepackage{appendix}
\usepackage{titlesec}
\usepackage{amssymb}
\usepackage{dcolumn}
\usepackage{bbm}
\usepackage{float}
\usepackage{amsmath}
\usepackage{bbold}
\usepackage{makecell}
\usepackage{tikz}
\usetikzlibrary{automata, positioning}
\usetikzlibrary{arrows}
\usepackage{multirow}
\usepackage{mathtools}
\usepackage{booktabs} 
\usepackage{siunitx}
\def\block(#1,#2)#3{\multicolumn{#2}{c}{\multirow{#1}{*}{$ #3 $}}}

\graphicspath{{./images/}}
\DeclareGraphicsExtensions{.pdf,.png}

\begin{document}
\title{Revealing Hidden States in Quantum Dot Array Dynamics: Quantum Polyspectra Versus Waiting Time Analysis}
\author{Markus Sifft}
\thanks{These authors contributed equally to this work.}
\address{Ruhr University Bochum, Faculty of Physics and Astronomy, Experimental Physics VI (AG), Universit\"atsstra{\ss}e 150, D-44780 Bochum, Germany}
\author{Johannes C. Bayer}
\thanks{These authors contributed equally to this work.}
\affiliation{Institut f\"ur Festk\"orperphysik, Leibniz Universit\"at Hannover, D-30167 Hanover, Germany}
\affiliation{Physikalisch-Technische Bundesanstalt, D-38116 Braunschweig, Germany}
\author{Daniel H\"agele}
\address{Ruhr University Bochum, Faculty of Physics and Astronomy, Experimental Physics VI (AG), Universit\"atsstra{\ss}e 150, D-44780 Bochum, Germany}
\author{Rolf J. Haug}
\affiliation{Institut f\"ur Festk\"orperphysik, Leibniz Universit\"at Hannover, D-30167 Hanover, Germany}

\date{\today}

\begin{abstract}
	Quantum dots (QDs) are pivotal for the development of quantum technologies, with applications ranging from single-photon sources for secure communication to quantum computing infrastructures. Understanding the electron dynamics within these QDs is essential for characterizing their properties and functionality. Here, we show how by virtue of the recently introduced quantum polyspectral analysis of transport measurements, the complex transport measurements of multi-electron QD systems can be analyzed. This method directly relates higher-order temporal correlations of a raw quantum point contact (QPC) current measurement to the Liouvillian of the measured quantum system. By applying this method to the two-level switching dynamics of a double QD system, we reveal a hidden third state, without relying on the identification of quantum jumps or prior assumptions about the number of involved quantum states. 
	We show that the statistics of the QPC current measurement can identically be described by different three-state Markov models, each with significantly different transition rates. 	
	Furthermore, we compare our method to a traditional analysis via waiting-time distributions for which we prove that the statistics of a three-state Markov model is fully described without multi-time waiting-time distributions even in the case of two level switching dynamics. Both methods yield the same parameters with a similar accuracy. The quantum polyspectra method, however, stays applicable in scenarios with low signal-to-noise, where the traditional full counting statistics falters. Our approach challenges previous assumptions and models, offering a more nuanced understanding of QD dynamics and paving the way for the optimization of quantum devices.
\end{abstract}

\maketitle

\section{Introduction}

Quantum dots (QDs) have become integral to the advancement of nanoelectronics and quantum technologies, serving as building blocks for a new generation of devices. These nanoscale structures are utilized in various fields, such as single-photon sources essential for secure communication in cryptography and advancements in photonic computing \cite{HeindelNJP2012, MuellerNatPho2014, GschreyNatComm2015, SenellartNNANO2017, TommNatNano2021, ZhaiNatNano2022}, as well as in quantum sensing where they offer high-sensitivity detection capabilities for fluorescence light \cite{ZhangBiosensors2023} and temperature measurements \cite{Boeyens_2023}. When organized into arrays, QDs provide a scalable infrastructure for quantum computing \cite{LossPRA1998, PettaScience2005, PhillipsNATURE2022}.

\begin{figure}[t]
	\centerline{\includegraphics[width=1\columnwidth]{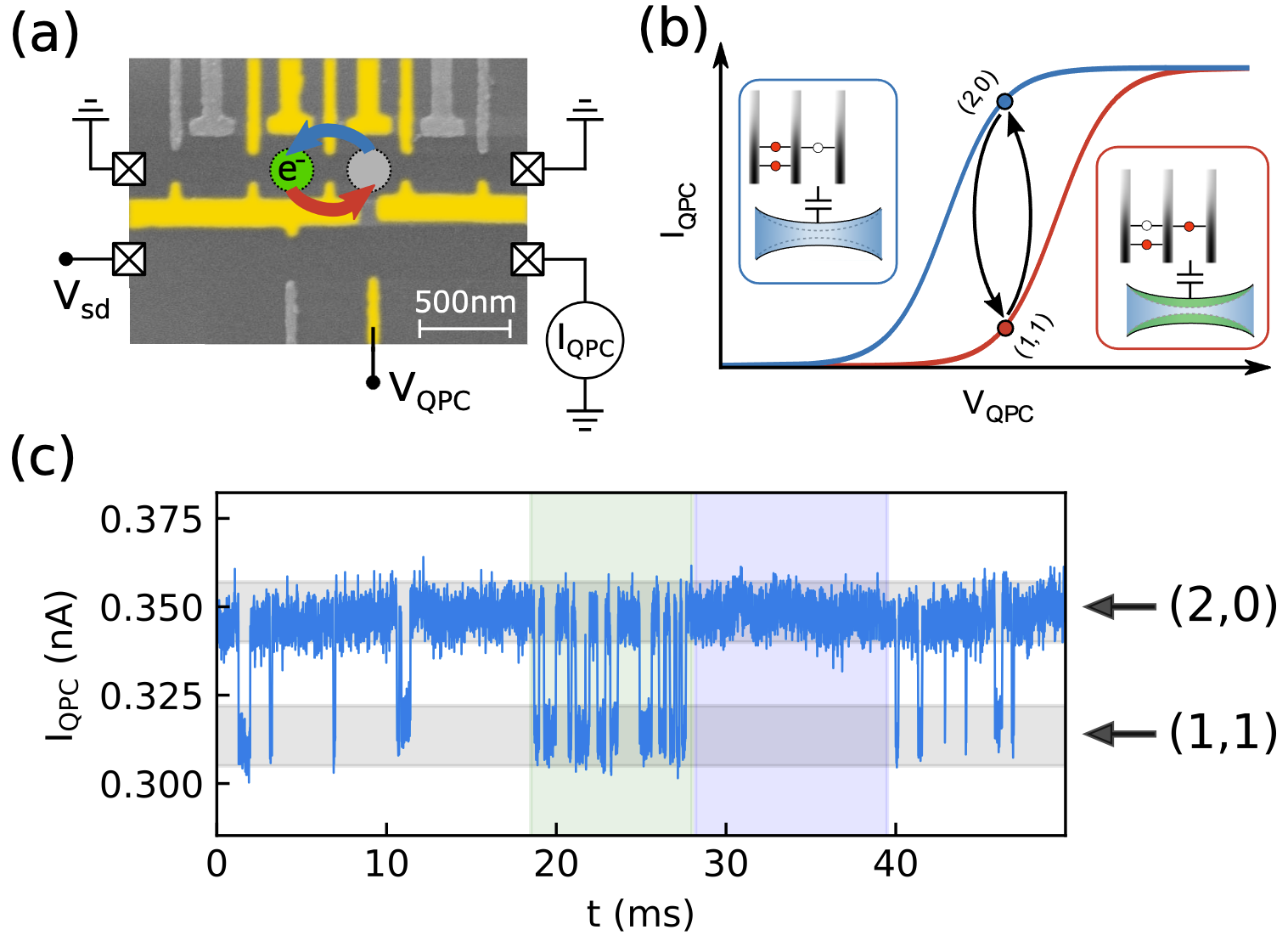}}
	\caption{(a) Scanning electron micrograph of the double quantum dot (QD) structure under investigation. The gates (yellow) deplete the underlying electron gas to form two distinct QDs (circles). Electron tunneling between dots, indicated by red and blue arrows, is detected through a quantum point contact (QPC) situated below the right QD by measuring the current $I_\textrm{QPC}$. (b) The resistance of the QPC is modulated by the presence of an electron in the right QD, providing information on its occupation state. (c) $I_\textrm{QPC}$ shows telegraph noise of the switching between two levels attributed to the charge configurations (2,0) and (1,1). Periods of rapid switching (green) and slower or no switching (blue) appear.}
	\label{fig:1}
\end{figure}

The functionality of QDs in these technologies is deeply rooted in the dynamics of the electrons they host. For instance, the performance of single-photon sources can be critically influenced by irregular electron tunneling that leads to blinking \cite{EfrosPRL1997, JahnPRB2015}. Moreover, the coherence of photon emission, vital for photon indistinguishability in quantum key distribution, is directly tied to the controlled dynamics of electron spin and interactions \cite{SantoriNature2002, WarburtonNatMat2013, SenellartNNANO2017}. Recently, it was shown in self-assembled QDs, that the coupling of electrons to host nuclei can be used to cool the latter leading to a strong increase in spin coherence times \cite{NguyenPRL2023}. In quantum computing, the controlled manipulation of electron spin dynamics within QDs forms the basis of qubit operations \cite{LossPRA1998, HansonRevModPhys2007}. Electron interactions within and between dots, along with the influence of the external environment, dictate qubit coherence and error rates, essential for the efficacy of quantum logic gates and robust entanglement over scalable QD arrays \cite{NowackScience2011, WatsonNature2018}. 

A QD device is collectively characterized by all aspects that influence the QD's temporal occupation dynamics, such as tunneling couplings, interactions both within a single dot and between different dots, as well as the effect of external environments. These properties can hence be accessed via measurements of the occupation dynamics. For example, in self-assembled double quantum dots it was shown that spin-coherence times can even be directly extracted from measurements of the tunneling current \cite{DaniCommPhys2022, DaniPRB2024}. These tunneling currents are often measured continuously via a quantum point contact (QPC) in the case of gate defined QD arrays \cite{ubbelohdeNATCOMM2012} or via resonance fluorescence in the case of self-assembled QDs \cite{kurzmannPRL2019}. In the former method, the strength of the probe current through the contact is related to the state of a QD, 
while in the latter method, it is determined by the fluorescence intensity. In the case of a strong coupling between the probe and the system or a high probe rate (\textit{strong measurement}), both methods reveal the state of the QD resulting in a detector output that exhibits telegraph noise due to quantum jumps in the occupation dynamics.
In the case of a weak coupling or a low probe rate (\textit{weak measurement}), however, the detector output is dominated by shot noise and quantum jumps are no longer visible \cite{sifftPRA2024}.

Measurement traces containing clear jumps are traditionally analyzed via the full counting statistics (FCS) \cite{levitovJMP1996, bagretsPRB2003}. Related quantities like cumulants \cite{flindtPNAS2009, cookPRA1981}, factorial cumulants \cite{kamblyPRB2011, stegmannPRB2015} and waiting-time distributions \cite{BrangeSciAdv2021, kleinherbersPRR2023}, or second- and third-order spectra of the frequency-resolved counting statistics \cite{emaryPRB2007} therefore rely on the identification of quantum jumps. Clearly, the FCS requires a high signal-to-noise ratio and a strong, continuous measurement that induces these jumps. However, these conditions are not always met, particularly in weak measurement regimes where quantum jumps are obscured by noise \cite{kungPRB2009}. Even recent advancements, which attempt to account for misclassifications in jump detection \cite{kerskiSR2023}, struggle in noisy weak measurement conditions. Furthermore, strong measurement conditions are not always desirable, as they suppress coherent dynamics — dynamics that are potentially very interesting, for example, when studying coherent electron and host nuclei spin dynamics \cite{NguyenPRL2023}. While there are examples in which noisy weak measurements on a single QD were evaluated via higher-order correlations up to third order \cite{kungPRB2009}, there is no straightforward way to generalize this method to larger systems.

Only recently, the analysis of continuous quantum measurements via so-called quantum polyspectra has been shown to fill this critical gap in traditional methodologies for evaluating QD dynamics from the strong to the weak measurement regime even under noisy conditions \cite{hagelePRB2018, sifftPRR2021}. Quantum polyspectra thus provide a solution to many issues discussed in a recent tutorial by Landi \textit{et al.} \cite{LandiPRX2024}. Our approach uses the raw output of the detector and compares its polyspectra — higher-order generalizations of the conventional power spectrum — with theoretical predictions derived from a device model. This comparison allows for finding model parameters, thereby characterizing the quantum system without the necessity of identifying individual quantum jumps. The device model is stated in terms of the stochastic master equation, an approach that accommodates general quantum systems under measurement and allows for the description of coherent quantum and incoherent dynamics as well as tuning of the measurement strength \cite{jacobsCP2006, korotkovPRB1999, korotkovPhysRevB2001, barchielliNC1982}. Under low-noise conditions, the quantum polyspectra method yields results consistent with those of the FCS and has even been demonstrated to be viable for the analysis of weak measurements and single-photon measurement on single QDs \cite{sifftPRR2021,SifftPRA2023,sifftPRA2024}.

Here, we apply quantum polyspectra to the direct analysis of more complex, continuous, strong measurements conducted on a gate-defined double QD system, as depicted in Figure \ref{fig:1}(a). The full state structure and dynamics of the system will be characterized based on a current measurement through a QPC [excerpt shown in Fig. \ref{fig:1}(c)] that is capacitively coupled to the QDs. Despite apparent binary switching between two levels, our analysis uncovers a hidden third state.
We demonstrate how a model of the system can be inferred directly from the data without any prior assumptions about state structures. Besides revealing an underlying three-state Markov model, we find that the model is not unique. This contrasts sharply with previous analyses of a similar system that assumed a physically motivated five-state model \cite{MaisiPRL2016}. Our method shows that assumption-free analysis of the measurement data does not conclusively require five underlying states. Therefore, our analysis approach being free from prior assumptions ensures that no potential explanations for the system dynamics are overlooked and is particularly promising for exploring systems with previously unknown dynamics.


\begin{figure}[b]
	\centerline{\includegraphics[width=1\columnwidth]{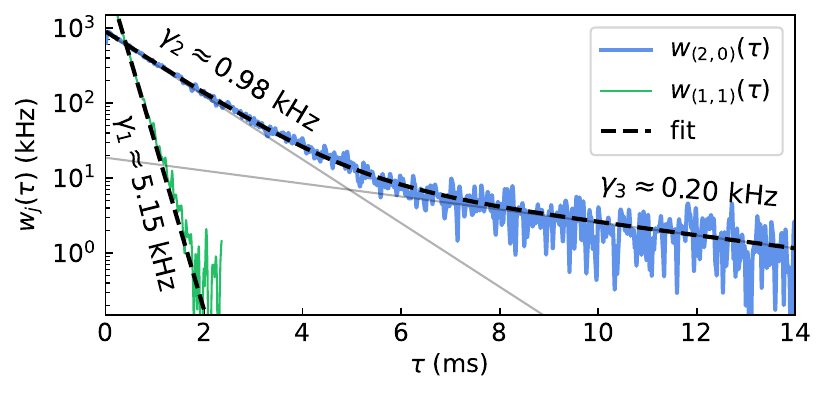}}
	\caption{Comparison of waiting-time distributions (WTDs) for electron configurations $(2,0)$ and $(1,1)$ in a double quantum dot. The WTD $w_{(2,0)}(\tau)$ represented by a thick blue line exhibits a bi-exponential decay, suggesting involvement of multiple states. In contrast, $w_{(1,1)}(\tau)$ (thin green line) decays mono-exponentially. Fits of the WTDs via Eqs.~(\ref{eq:general_wtd1}) and (\ref{eq:general_wtd2}) are depicted as dashed lines along with the resulting decay rates.}
	\label{fig:WTDs}
\end{figure}


\section{Quantum polyspectra of occupation dynamics}
\label{sec:device}

\begin{figure*}[t]
	\centerline{\includegraphics[width=\textwidth]{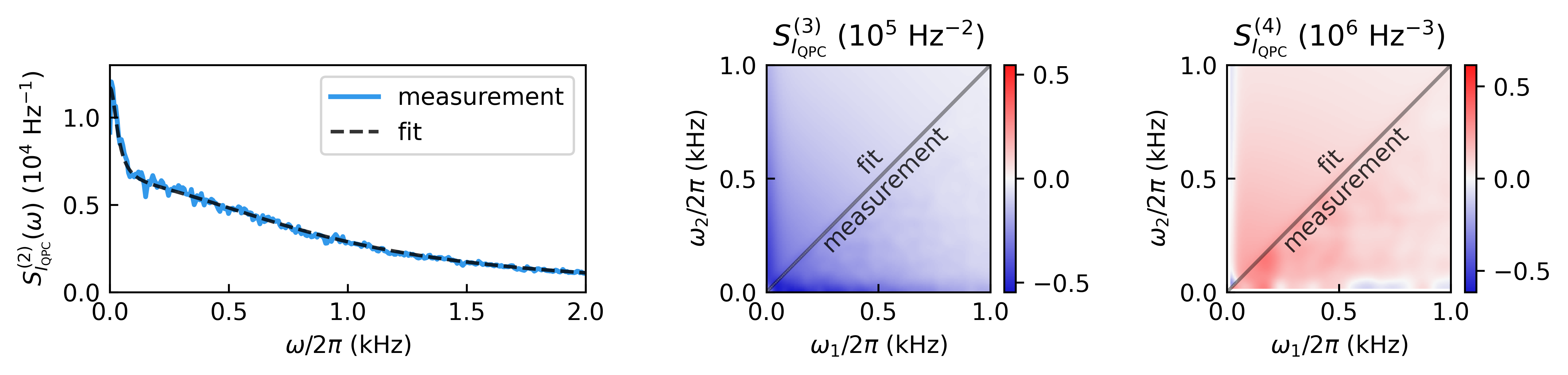}}
	\caption{Comparison of experimental polyspectra $S^{(2)}_{I_\textrm{QPC}}$, $S^{(3)}_{I_\textrm{QPC}}$, and $S^{(4)}_{I_\textrm{QPC}}$ calculated from the QPC current $I_\text{QPC}(t)$ (measurement) and their analytic counterparts derived from a three-state Markov model with two output levels (fit). The model spectra show excellent agreement with all experimental spectra. Notably, each spectrum displays both sharp and broad peak features, which excludes a description of random telegraph noise by just a simple two-state Markov model.}
	\label{fig:2}
\end{figure*}

The device we are going to study is a Schottky gate defined double QD based on a GaAs/AlGaAs heterostructure forming a two-dimensional electron gas approximately $\SI{110}{nm}$ below the surface, similar to the device used in \cite{BayerAnnPhys2019}. The double QD is formed electrostatically in the electron gas by applying negative voltages to the gates. The plunger gate voltages hereby provide tuning parameters for the energy levels of the QDs, the tunnel barrier gates allow to manipulate the coupling to the electron reservoirs as well as between the two QDs. A galvanically separated QPC in the vicinity of the right QD is tuned into a regime of high sensitivity to local potential changes and is used as a highly sensitive charge detector. The sensitivity of the QPC charge detector allows to resolve single electron tunneling events in the double QD system and is used for the time resolved detection of charging events. In order to achieve a sufficient time-resolution and a high sensitivity, the QPC drain was connected to a low capacitance ($\SI{30}{pF}$) line and a low-noise \textit{FEMTO} LCA-100K-50M current amplifier. The time resolved data was acquired using an ADwin-Pro II real-time controller with a sampling rate of $\SI{400}{\kilo \hertz}$. All measurement were performed at temperatures of $T = \SI{1.5}{\kelvin}$. 

The double QD is set up to be an isolated double QD system \cite{BayerAnnPhys2019} charged with exactly two electrons via a controlled loading procedure. In a first step, the right QD is hereby fully depleted, i.e., the lowest level of the QD lies above the reservoir potentials. By applying highly negative voltages to the right tunnel barrier gate, the double QD is then isolated from the right electron reservoir. The left plunger gate is then used to fine tune the electron number in the left QD before using a ramp of the left tunnel barrier gate voltage to also decouple the double QD from the left reservoir and thus effectively isolating the double QD with a fixed number of electrons. The number of electrons in the left QD then defines the total number of electrons trapped inside the double QD. A voltage ramp at the left tunnel barrier gate with a rate of $\SI{0.5}{V/s}$ was used to initialize the system before starting the
time resolved measurements. The endpoint of the ramp was chosen such, that the cross coupling of the tunnel barrier gate on the energy levels tunes the system close to the degeneracy between the $(2,0)$ and the $(1,1)$ configuration, where $(N_L, N_R)$ are the electron numbers in the left and right QD.

Figure \ref{fig:1}(c) displays a segment of the charge detector current $I_\textrm{QPC}(t)$. A lower current indicates a more negative potential at the detector, suggesting an electron occupation close to the detector, which is situated in the vicinity of the right QD. Consequently, we associate the lower and higher current levels with the $(1,1)$ and $(2,0)$ charge configurations, respectively. The data also shows periods of rapid transitions (highlighted in green) and periods without transitions (highlighted in blue). A preliminary analysis of the dynamics is performed via the two waiting-time distributions (WTDs) $w_{(2,0)}(\tau)$ and $w_{(1,1)}(\tau)$, where $\tau$ are the times the system resides in the corresponding state [see Fig. \ref{fig:WTDs}]. The $(2,0)$ configuration exhibits a double exponential decay, while the WTD for the $(1,1)$ configuration is mono-exponential. This indicates that the system cannot be treated as a simple two-state model where single-exponential WTDs are expected.

In the remainder of this paper, we will compare conclusions drawn from the WTDs with results obtained from the quantum polyspectra approach. Please note, that unlike the WTD method, which relies on the identification of quantum jumps, the quantum polyspectra approach is directly based on the charge detector current, $I_\textrm{qpc}(t)$. Polyspectra of a stochastic quantity $z(t)$ have been defined by Brillinger in 1965 \cite{brillingerAMS1965}. They generalize the usual second-order power spectrum $S_z^{(2)}(\omega) \propto \langle z(\omega) z^*(\omega) \rangle$ to spectra that are of higher order of $z(\omega)$, where $z(\omega) = \int z(t) e^{i \omega t}\,dt$ is the Fourier transform of $z(t)$.
The definition of polyspectra
\begin{align}
	2\pi \delta(\omega_1+...+\omega_n)S_z^{(n)}&(\omega_1,...,\omega_{n-1})\nonumber\\ 
	&= C_n(z(\omega_1),..., z(\omega_n)) \label{eq:defPolyspectra}
\end{align} 
is based on the $n$th-order cumulant $C_n$, where
\begin{align}
	C_1(x) & = \langle x  \rangle  \nonumber \\
	C_2(x,y) & =  \langle x y \rangle - \langle x \rangle \langle  y \rangle \nonumber \\
	C_3(x,y,z) & =  \langle (x - \langle x \rangle) ( y - \langle y \rangle)  (z - \langle z \rangle) \rangle. 
\end{align}
The fourth-order cumulant can, e.g., be found in Refs. \cite{gardinerBOOK2009,hagelePRB2018}. The practical calculation of experimental polyspectra from a finite amount of data using unbiased cumulant estimators and the fast Fourier transformation is described in our Ref. \cite{sifftPRR2021}.
All experimental polyspectra in the article have been evaluated with our SignalSnap software library \cite{sifftSIGNALSNAP2022}.
The library also provides estimates for the errors of the spectral values. The errors enter the fitting procedure of model spectra discussed below [see App. \ref{app:ErrorEstimation}].

Figure \ref{fig:2} shows the second-, third-, and fourth-order spectrum of a measurement trace with an overall temporal length of 89~s. Background spectra had been subtracted that were recorded with the same system gate settings but with only a single electron loaded which occupied the stationary $(1,0)$ configuration. In the single electron setting, telegraph noise is absent but slow fluctuations and environmental influences persist, whose spectra need to be subtracted from spectra of recorded overall dynamics. After subtraction, the polyspectra in Figure \ref{fig:2} primarily reflect the true occupation dynamics. 

The spectra shown in Fig. \ref{fig:2} are clearly distinct from spectra of a simple two-state system. A two-state system exhibits a single Lorentzian peak in the second-order spectrum, whose width is given by the sum of the two tunneling rates \cite{sifftPRR2021}. The third-order spectrum changes sign with the difference of the tunneling rates and is required for separating the tunneling rates. Including the fourth-order spectrum into the evaluation procedure further reduces the error of the retrieved system parameters.
In contrast to a simple two-state system, the second-order spectrum of our system exhibits a superposition of a broad and a narrow Lorentzian peak, both centered at zero frequency. Similarly, the third and fourth-order spectrum exhibit narrow features close to zero frequency. These spectra can be viewed as a \textit{fingerprint} of the system dynamics for which a model system is desired. Model parameters will be found below by fitting model spectra to their experimental counterparts.
Polyspectra of fifth or higher order have, to the best of our knowledge, not been used in literature, probably due to the computational demands and the increasing noise observed in cumulants of increasing order \cite{schefczikARXIV2019}.


\section{Modeling incoherent multi-state quantum dynamics}
In the following we model the behavior of the double QD system which is constantly monitored by a QPC. The transition dynamics between quantum states is described by a Liouvillian within a master equation approach. The measurement and its stochastic backaction on the system are included in the model by generalizing the master equation to the so-called stochastic master equation (SME) of continuous measurement theory \cite{jacobsCP2006}. General expressions for polyspectra of the SME detector output up to fourth order had been derived by Hägele and Schefczik in 2018 and will be used in the next section for analyzing experimental data \cite{hagelePRB2018}. 

The model system consists of $N$ quantum states $|j \rangle$. In case of absent quantum coherence of the states, the system can be described by a diagonal density matrix
$\rho(t) = \sum_j \rho_{jj} |j\rangle \langle j|$ where $\rho_{jj}$ is the probability of the system to be in state $j$. The dynamics of the system is given by the transition rates between pairs of states. For instance, an incoherent transition from state $|1\rangle$ to state $|2\rangle$ is represented by a jump operator $d = |2\rangle \langle 1|$. The equation of motion for the density matrix, considering a single transition, is then given by

\begin{equation}
	\dot{\rho} = \gamma {\cal D}[d](\rho),
\end{equation} 
where $\gamma$ is the transition rate and
\begin{equation}
	{\cal D}[d](\rho) = d \rho d^\dagger - (d^\dagger d \rho + \rho d^\dagger d) /2,
\end{equation} 
is a superoperator acting on the density matrix \cite{tilloyPRA2018}. To define an arbitrary $n$-state model, we have to set $n(n-1)$ different transition rates (a jump from one state to itself does not need to be defined). When denoting the jump operators as $d_{ij} = |j\rangle \langle i|$ we find the general master equation
\begin{eqnarray}
	\label{eq:master_eq}
	\dot{\rho} &=& \sum_{i \neq j}\gamma_{ij} {\cal D}[d_{ij}](\rho) \nonumber\\
	&=& \mathcal{L}_0(\rho),
\end{eqnarray} 
where we have introduced the Liouvillian $\mathcal{L}_0$ that contains all the terms on the right-hand side of the equation. Not all state pairs need to share jump operators; thus, some transition rates $\gamma_{ij}$ may be zero. 
In our specific case of describing tunneling dynamics, the density matrix $\rho(t)$ remains diagonal throughout the evolution, unlike in general quantum systems where coherences between states can result in non-zero off-diagonal elements of $\rho(t)$. The behavior of the master equation is for absent quantum coherence identical with that of a continuous Markov model. Below, we will therefore identify the states 
$| j \rangle$ often as Markov states rather than quantum states. However, if needed, coherent quantum dynamics given by a Hamiltonian $H$ could be introduced by adding
$-i(H \rho - \rho H) /\hbar$ to the Liouvillian.

The master equation [Eq.~\eqref{eq:master_eq}] yields for long times the equilibrium state $\rho_0 = \sum_j p_j |j\rangle \langle j|$ with the probabilities 
$p_j$ of finding the system in a certain state $| j \rangle$.
It does, however, not reproduce the actual stochastic behavior of the system under measurement and can not be used to simulate measurement traces $I_\textrm{QPC}(t)$, such as the telegraph noise shown in Fig. \ref{fig:1}(c). These limitations are overcome by the SME, where a measurement operator $A$ and a measurement strength $\beta$ is introduced to effectively describe both the stochastic system dynamics and the measurement outcomes \cite{jacobsCP2006,barchielliNC1982}. In our case of the QPC measurement, the operator $A$ needs to reflect the distinct detector output levels, $I_{\text{high}}$ and $I_{\text{low}}$.

For instance, for a three-state model, a correct choice for the measurement operator is
\begin{equation}
	A = I_\textrm{low} |0\rangle \langle 0| + I_\textrm{high}(|1\rangle \langle 1|+|2\rangle \langle 2|). \label{operatorA}
\end{equation}
All three-state models can be constructed using this operator and adapting the state labeling.

The relationship between the measured current $I_\text{QPC}(t)$ and the theoretical model is given by \cite{korotkovPhysRevB2001,sifftPRR2021}
\begin{equation}
	z(t) = \beta^2 I_\textrm{QPC}(t) = \beta^2 {\rm Tr}[\rho(t) (A + A^\dagger)/2]+ \beta \Gamma(t)/2, \label{SME_detector}
\end{equation}
where $z(t)$ represent the detector output, $\beta^2$ is the measurement strength, and $\Gamma(t)$ is the white background noise, characterized by $\langle \Gamma(t) \Gamma(t') \rangle = \delta(t-t')$. The equation reflects the fact that a measurement of $A$ yields information about the state of the system, albeit partially obscured by noise.
In our case, the switching behavior is obtained in the strong measurement limit $\beta \gg 1$. This regime ensures that the state of the system is promptly revealed, corresponding to the sharp transitions as typically observed in the telegraph noise of $I_{\text{QPC}}(t)$.

The SME governs the behavior of the system during measurement, expressed as follows (adopting the notation from \cite{sifftPRR2021}):
\begin{eqnarray}
	d \rho & = & {\cal L}_0\rho \, dt + \beta^2 {\cal D}[A](\rho) \, dt \nonumber \\
	& & + \beta(A \rho + \rho A^\dagger - {\rm Tr}[(A + A^\dagger) \rho ] \rho)\,dW, \label{eq:smeV2}
\end{eqnarray}
where the last line describes a stochastic measurement backaction driven by the Wiener-process $W$, where formally $\dot{W} = \Gamma(t)$. This term, along with the modification in the first line, describes how the quantum system experiences a collapse into an eigenstate of $A+A^\dagger$, with the rate of collapse scaling with $\beta^2$. 

\begin{figure*}[t]
	\centerline{\includegraphics[width=15.5cm]{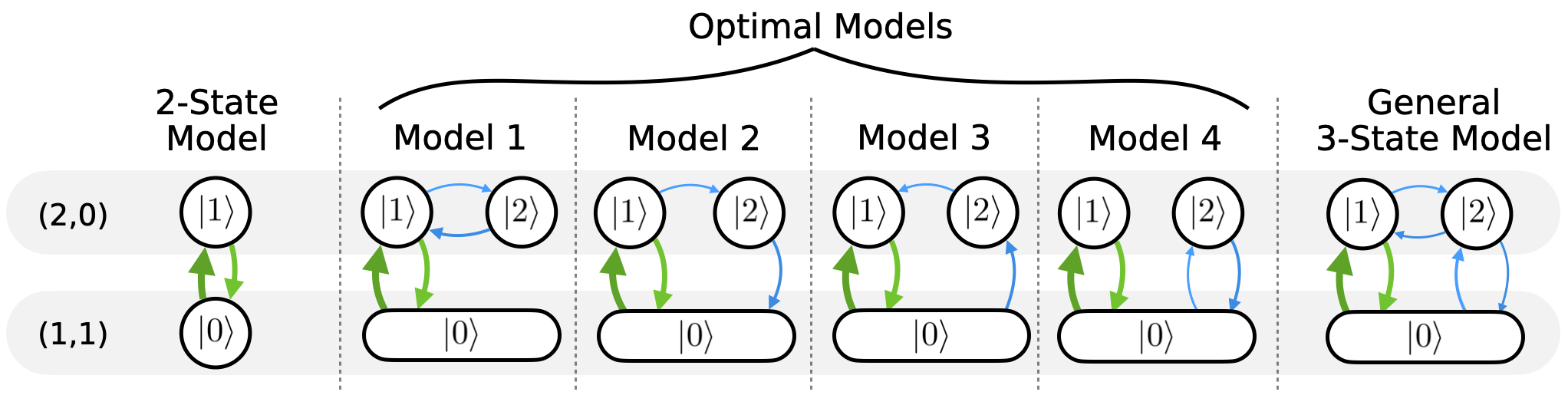}}
	\caption{Model candidates for double dot charging dynamics: The comparison of experimental data with theoretical quantum polyspectra suggests that the $(2,0)$ state is represented by two Markov states ($|1 \rangle$ and $|2 \rangle$), whereas the $(1,1)$ state corresponds to a single state $|0 \rangle$. Thicker arrows represent larger rates between the Markov states. Fast switching dynamics, highlighted in green in Figure \ref{fig:1}, occur between $|0 \rangle$ and $|1 \rangle$. Slow switching periods, shown in blue in Figure \ref{fig:1}, are associated with state $|2 \rangle$, which is linked to other states by low transition rates, leading to long constant signals associated with the $(2,0)$ configuration. We show in App. \ref{app:TheoreticalWTD} that there is no unique three-state Markov model for describing the measured telegraph noise signal.}
	\label{fig:model_comparison}
\end{figure*}

The comparison of a measured stochastic detector current and the model current $z(t)$ needs to be done in terms of their statistical properties. Here, the experimental polyspectra of the detector current will be compared with model spectra. Quantum mechanical expressions for such quantum polyspectra follow without any approximation from equations Eq. \eqref{SME_detector} and Eq. \eqref{eq:smeV2} \cite{hagelePRB2018}. They are given in terms of the propagator
\begin{equation}
	{\cal G}(\tau) = e^{{\cal L}\tau}\Theta(\tau),
\end{equation} 
where ${\cal L} = {\cal L}_0 + \beta^2 {\cal D}[A](\rho)$ includes the effects of measurement-induced damping and $\Theta(\tau)$ is the Heaviside-step-function.
The steady state is
\begin{equation}
	\rho_0 = {\cal G}(\infty) \rho(t)
\end{equation} 
and the measurement superoperator is 
\begin{equation}
	{\cal A} x = (A x + x A^\dagger)/2.
\end{equation}
Very compact expressions for the quantum polyspectra follow after introducing the modified propagator
${\cal G}'(\tau) = {\cal G}(\tau)- {\cal G}(\infty)\Theta(\tau)$ and the modified measurement operator ${\cal A}' x = {\cal A} x - {\rm Tr}({\cal A}\rho_0)x$. As per \cite{hagelePRB2018}, the first-order spectrum
\begin{eqnarray}
	S_{z}^{(1)} &=& \langle z(t) \rangle \nonumber \\
	& = & \beta^2 {\rm Tr}[{\cal A}\rho_0] 	\label{eq:S1},
\end{eqnarray}
is essentially the expectation value of the measurement operator.
The second-order spectrum is
\begin{eqnarray}
	S_{z}^{(2)}(\omega) &=& \beta^4 ( {\rm Tr}[{\cal A}'{\cal G}'(\omega){\cal A}'\rho_0] + {\rm Tr}[{\cal A}'{\cal G}'(-\omega){\cal A}'\rho_0] ) \nonumber \\
	& & \hspace{-0.3cm} + \beta^2/4, 
	\label{eq:S2}
\end{eqnarray}
where ${\cal G}'(\omega) = \int {\cal G}'(\tau) e^{i \omega \tau}\,d\tau$ is the Fourier transform of ${\cal G}'(\tau)$.
The third-order spectrum is
\begin{eqnarray}		
	S_{z}^{\rm (3)}(\omega_1,\omega_2,\omega_3 = -\omega_1-\omega_2) &= & \nonumber \\
	& & \hspace{-40mm}	\beta^6\hspace{-6mm} \sum_{\{k,l,m\} \in \text{prm.} \{1,2,3\}} \hspace{-6mm} {\rm Tr}[{\cal A}'{\cal G}'(\omega_m){\cal A}'{\cal G}'(\omega_m + \omega_l){\cal A}'\rho_0], \label{eq:S3}		
\end{eqnarray}
where the sum regards all six permutations (prm.) of the indices of the $\omega_j$s \cite{footnote1}. The fourth-order spectrum is given in App. \ref{app:QuantumPolyspectra}. The calculation of spectra from ${\cal L}$ and ${\cal A}$ and fitting to experimental spectra are performed via our QuantumCatch library 
(general quantum case) or our MarkovAnalyzer library (optimized code for the pure Markov case)
which are based on the ArrayFire library \cite{sifftQUANTUMCATCH2023,sifftMarkovAnalyzer2023,Yalamanchili2015}.
\section{Analysis of the double QD charging dynamics}
Our goal is to characterize the double QD system introduced above by determining the number of states involved in the dynamics of the system and their corresponding transition rates. This is achieved without making prior assumptions about the system size or state connectivity (number of involved jump operators). In the purely data-driven approach, we seek an \textit{optimal} system model that accurately describes the system dynamics with the least parameters possible. This idea is well established in the fields of biology, physiology, and data science and will here be applied to 
the characterization of a quantum system.

Typically, the model selection process is guided by a so-called information criterion,
such as the Akaike Information Criterion (AIC) \cite{AkaikeIEEE1974}. The AIC weights the accuracy of the fit against the complexity of the underlying model to prevent overfitting. This results in a single number score $AIC = 2k - 2\ln(RSS/n)$, where $k$ is the number of parameters, $n$ is the number of points in the dataset (total number of spectral values in our case), and $RSS/n$ represents the mean squared error derived from the residual sum of squares (RSS). The model with the lowest AIC is 
more likely to capture the true system behavior. 
The AIC values can shift if the error squares are scaled by a constant factor (like in the case of changing units). This shift is, however, the same for different models. Hence, the quality of different models is given rather by the difference of their AIC values than by their ratio. Large AIC numbers with small differences between models as in the case below, therefore, should not lead to the wrong conclusion that these models are almost equally useful.

\begin{table*}[t]
	\caption{The table lists the transition rates that were obtained from fitting either the quantum polyspectra or waiting-time distributions 
		of different Markov models to the corresponding experimental quantities of the same measurement trace. Across Models 1 to 4, both quantum polyspectra (QPS) and WTD methods yield consistent results within error margins of three standard deviations. The fitting errors have been determined from fitting 250 simulated measurements (transition rates taken from the experiment). Notably, quantum polyspectra show improved error precision particularly for the rate $\gamma_{21}$ while the WTDs yield for Model 1 and Model 2 a smaller error for $\gamma_{10}$. Please note, that the general three-state model cannot converge to a unique set of transition rates, as it already contains Model 1 to Model 4 as special cases. Therefore, no errors can be given.
	}
	\centering
	\sisetup{table-format=4.0, group-four-digits=true}
	\begin{tabular}{cccccccccccc}
		\toprule
		\multirow{2}{*}{\textbf{Rates / Hz}} & \multirow{2}{*}{\makecell{\textbf{2-State}\\\textbf{Model}}} & \multicolumn{2}{c}{\textbf{Model 1}} & \multicolumn{2}{c}{\textbf{Model 2}} & \multicolumn{2}{c}{\textbf{Model 3}}& \multicolumn{2}{c}{\textbf{Model 4}}& \multirow{2}{*}{\makecell{\textbf{General}\\\textbf{3-State Model}}} \\
		\cmidrule(lr){3-4} \cmidrule(lr){5-6} \cmidrule(lr){7-8} \cmidrule(lr){9-10}
		& & \textbf{QPS} & \textbf{WTD} & \textbf{QPS} & \textbf{WTD} & \textbf{QPS} & \textbf{WTD} & \textbf{QPS} & \textbf{WTD} & \\
		\midrule
		$\gamma_{01}$ &$5267 $ & $5115 \pm 111$   & $5166 \pm 76$  & $5115 \pm 110$       & $5166 \pm 76$  &   $4783 \pm 121$    & $4780 \pm 129$  & $4715 \pm 132$  & $4682 \pm 184$     &$4785$\\
		$\gamma_{02}$ & -           & -           & -              & -                  & -             &      $332 \pm 74$    & $387 \pm 106$    & $400 \pm 94$   & $484 \pm 169$             &$315$\\
		$\gamma_{10}$ &$ 405 $  & $953 \pm 40$    & $904 \pm 19$  & $953 \pm 40$         & $904 \pm 19$   & $1020 \pm 52$       & $977 \pm 41$    & $1020 \pm 52$     & $977 \pm 42$           &$990$ \\
		$\gamma_{12}$ & -       & $54 \pm 13$     & $57 \pm 15$    & $67 \pm 18$     & $73 \pm 25$    & -                 & -               & -               & -              &$5.8$ \\
		$\gamma_{20}$ & -          & -            & -               & $173 \pm 36$   & $196 \pm 85$   & -                 & -              & $173 \pm  35$ & $196 \pm 87$     &$165$ \\
		$\gamma_{21}$ & -          & $185 \pm 41$ & $212 \pm 94$  & -                    & -             &   $173 \pm 36$     & $196 \pm 85$   & -                & -        &$0.2$  \\ \hline
		\makecell{AIC\\ (QPS only)}      & 976842  & 779779 & -	 & 779779 & -         & 779779 & -        & 779779 & -              &779783\\
		\bottomrule
	\end{tabular}
	\label{tab:rates}
\end{table*}

Multiple models were tested to find the optimal model with minimal $AIC$ for our system. Our detailed comparison focuses on one two-state model, four three-state models with reduced connectivity, and a general three-state model, as illustrated in Figure \ref{fig:model_comparison}. The respective transition rates together with fitting errors and AIC values are given in Table \ref{tab:rates} for a selection of models. A simple two-state model only yields either the fast or the slow switching dynamics separately. Depending on the start values of the transition rates, such a fit settles for either of the two sub-dynamics. The table displays the results of the two-level fit that recovered fast dynamics. While the transition rates fall into the correct order of magnitude, a comparison of the $AIC$ values with those from the three-state systems immediately reveals that the dynamics is not well approximated by a two-state system. 

The same minimal $AIC$ was found for four different models which comprise a four parameter subset of the general six-parameter three-state model and are shown in Figure \ref{fig:model_comparison}. These four transitions are just enough to capture both the fast and the slow dynamics of the system. A physical interpretation of these models follows below. The general three-state system results in an equally precise fit, however, while introducing two unnecessary parameters immediately visible in the AIC value, which increases by four. Larger models result in even higher AIC values (not shown). 

We find a number of properties of continuous three-state Markov models with only two output levels that to the best of our knowledge have not been shown before.
First, the waiting-time distributions have always the form
\begin{eqnarray}
	\label{eq:general_wtd1}
	w_{\textrm{low}}(\tau) &\propto& e^{- \gamma_1 \tau}, \\ \label{eq:general_wtd2}
	w_{\textrm{high}}(\tau) &\propto& e^{- \gamma_2 \tau} + \alpha e^{- \gamma_3 \tau},
\end{eqnarray}
with three distinct decay constants $\gamma_i$ and a weighting factor $\alpha$. The mono-exponential WTD $w_{\textrm{low}}(\tau)$ belongs 
to the low output level that represents a unique Markov state [in our case the electron configuration (1,1)].
The double-exponential WTD $w_{\textrm{high}}(\tau)$ belongs to the output level corresponding to the two other Markov states [different spin orientations of configuration (2,0)].
Second, all higher-order waiting-time distributions can be expressed in terms of the first-order WTDs $w_{\textrm{low}}(\tau)$ and $w_{\textrm{high}}(\tau)$ [see App. \ref{app:ProofHigherOrder}].
Third, three-state Markov models that result in WTDs $w_{\textrm{low}}(\tau)$ and $w_{\textrm{high}}(\tau)$ are not unique. We proof for the four optimal
three-state Markov models with only four non-zero transition rates that they share identical statistics of their output [see App. \ref{app:TheoreticalWTD}]. This implies also that polyspectra for different models are identical, which explains our result of the same AIC-values. The three properties of three-state Markov models cannot be generalized to four-state Markov models or periodically driven systems \cite{BayerArxiv2024}!

Table \ref{tab:rates} compares the transition rates found for the optimal models (Model 1 to 4) applying the quantum polyspectra method and an analysis via the WTDs. The latter was performed by predicting the full WTDs (Eqs.~\eqref{eq:wtd1} and \eqref{eq:wtd2}) and comparing them to their estimation from $I_\textrm{QPC}(t)$. Both methods yield consistent results within the error margins of three standard deviations. The errors have been estimated by performing fits on 250 simulations. The simulations mimicked the telegraph noise and a similar white background noise level as in the actual experiment. The standard deviation of the distribution of fitting results for the transition rates was used to estimate the error of the evaluation methods. The fact that the quantum polyspectra and the WTDs methods do not yield exactly the same values and errors should not be surprising, given the distinct types of information each method accesses. The WTD method solely captures the intervals between quantum jumps (i.e., periods of non-activity), thus providing a more limited view of the dynamics of the system. In contrast, the quantum polyspectra method utilizes correlation functions at one, two, three, and four time points, offering a significantly richer 
fingerprint of the data that captures much more dynamic behavior than what is revealed by the WTDs alone.

Despite the richer information set provided by the quantum polyspectra, not all parameters are determined with greater accuracy compared to those derived from the WTDs. 
This limitation can be attributed to the computational constraints of the quantum polyspectra analysis. Specifically, the fitting procedure was limited to a maximum frequency of 5 kHz with a resolution of 7.5 Hz, which was a compromise for managing the computational load. Extending the frequency range might have enhanced the precision of the parameter estimates, but were not pursued in this study.

Moreover, while significant information may reside in the full three-dimensional spectrum $S^{(4)}(\omega_1, \omega_2, \omega_3)$, we utilized only the reduced form $S^{(4)}(\omega_1, \omega_2, -\omega_1)$ due to computational constraints. 
It is also important to note that the experimental polyspectra may exhibit additional contributions from environmental noise. They are not regarded in the error estimates that were determined from simulations that did not regard environmental noise apart from a general white background noise. 


\section{Discussion}
\begin{figure}[b]
	\centerline{\includegraphics[width=\columnwidth]{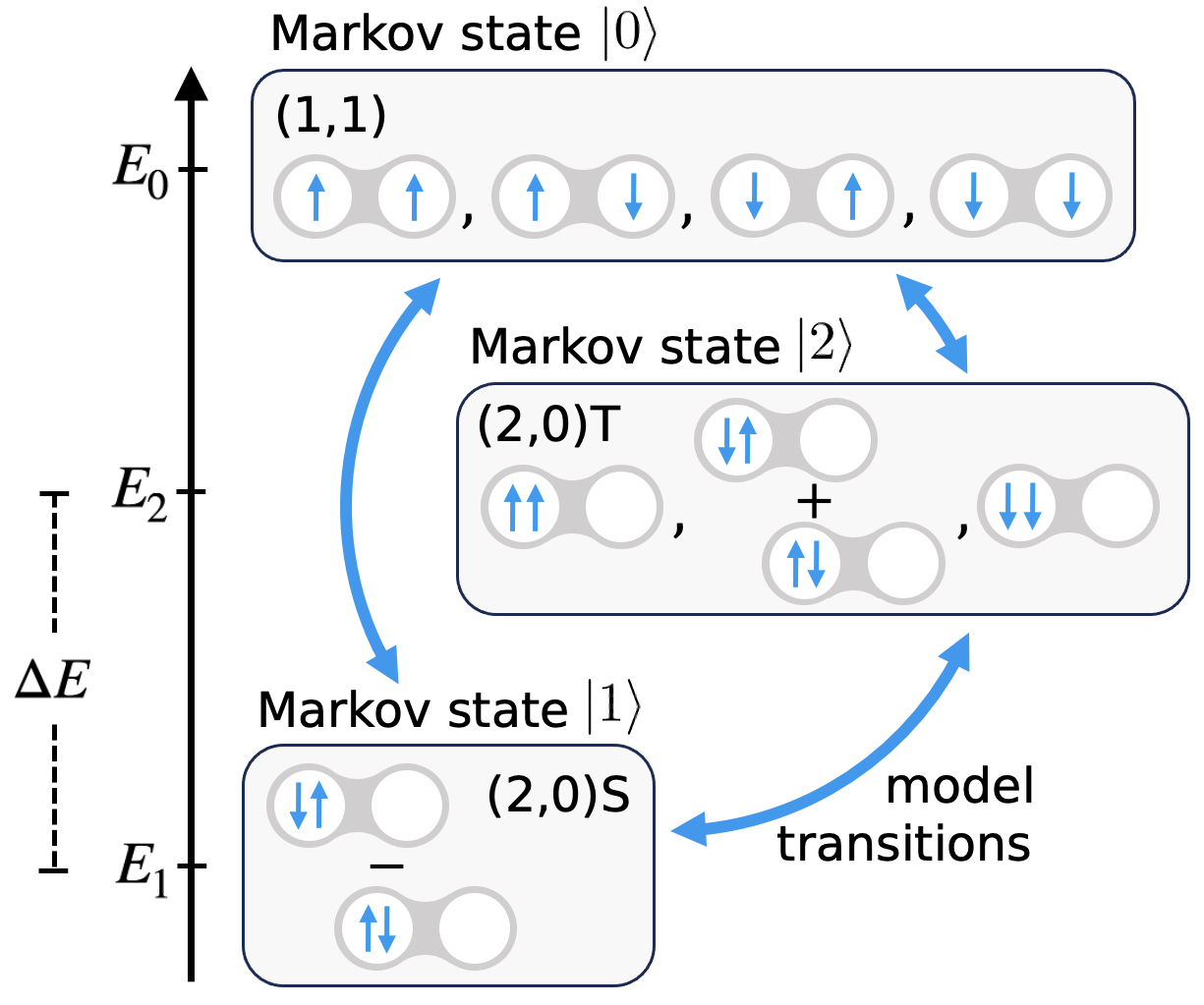}}
	\caption{Physical interpretation of the Markov states within the double QD model.
		The schematic illustrates the three Markov states, which represent multiple underlying quantum states depicted as single spins located in the left or right QD. These Markov states are ordered by their relative energies, as determined by their steady-state probabilities. Markov state $|1\rangle$ is identified as the lowest energy singlet state in the $(2,0)$ configuration. Markov state $|2\rangle$ is associated with the triplet states in the $(2,0)$ configuration, and Markov state $|0\rangle$ corresponds to all $(1,1)$ configuration (see main text).}
	\label{fig:physical_interpretation}
\end{figure}

Next, we give a physical interpretation of the three-state Markov models in terms of the two-electron charge configurations (2,0) and (1,1) taking into account also their spin configurations. 
In the previous section, we showed that three Markov states are sufficient to describe the measurement traces, while we have to consider here a total of eight quantum states as the two charge configurations come with four spin-configurations each. We argue that different spin-configurations can correspond to the same Markov state. 
The QD is tuned close to the degeneracy point of the $(2,0)$ and $(1,1)$ charge configurations. 
Since the $(1,1)$ configuration corresponds to the low current value $I_\textrm{low}$ in the measurement, we identify it with the Markov state $|0\rangle$ which was assigned the value $I_\textrm{low}$ in the measurement operator $A$, see Eq.~(\ref{operatorA}). The Markov states $|1\rangle$ and $|2\rangle$ correspond to $I_\textrm{high}$. A physical interpretation must regard the steady state probabilities of the Markov states for the Markov models, which are $(\rho_{00}, \rho_{11}, \rho_{22}) = (0.126, 0.676, 0.197)$ for the first model and $(0.126,0.633,0.241)$, $(0.126, 0.632 ,0.242)$, and $(0.126, 0.583,0.291)$ respectively for the three other models. Consequently, Markov state $|1\rangle$ is the candidate for quantum states with the highest probability and therefore with the lowest energy.
In Figure \ref{fig:physical_interpretation} we suggest an energy scheme for all quantum states and their corresponding Markov states which is consistent with the values for $(\rho_{00}, \rho_{11}, \rho_{22})$ for all four models. Markov state $|1\rangle$ is identified as the lowest energy state which is in the configuration (2,0) with both electrons in the left quantum dot being in the favorable singlet state. At slightly higher energy, the three-fold degenerate triplet states in configuration (2,0) are displayed and associated with the Markov state $|2\rangle$. The energetically least favorable states correspond to spatially separate electrons [configuration (1,1)] with an arbitrary four-fold degenerate spin configuration due to a negligible exchange interaction. These quantum states belong to the Markov state $|0\rangle$.
The singlet-triplet splitting in the (2,0) configuration follows from $\rho_{22}/\rho_{11} = 3 \exp(-\Delta E/ k_{\rm B} T)$ with $k_{\rm B} T \approx \SI{130}{\micro eV}$ (corresponding to 1.5~K) as about $\Delta E \approx \SI{233}{\micro eV}$ for Model 4 which is a typical value for gate-defined GaAs quantum dots. For Model 1 we find $\Delta E \approx \SI{303}{\micro eV}$ which is also a reasonable value. The values for Model 2 and 3 are comparable to those of Model 1 and 4.
Our interpretation of a Markov state representing several quantum states implies that these quantum states share effectively the same transition rates towards quantum states of other Markov states. A transition from triplet-states of $|2\rangle$ to the singlet-state $|1\rangle$ requires a spin flip, independent of the specific initial triplet state. We therefore argue that the transition rates between quantum states of $|2\rangle$ and $|1\rangle$ are always the same. In case of transitions from quantum states of the Markov state $|0\rangle$, one might argue that the state $\left |\uparrow \uparrow\right\rangle$ tunnels fast to the triplet state $\left |\uparrow \uparrow\right\rangle$ of $|2\rangle$ due to an identical spin configuration.
A state $\left |\uparrow \downarrow\right\rangle$ on contrary may tunnel slower due to an unequal spin-configuration with all triplet-states of $|2\rangle$. We argue, however, that fast scattering of spins within the Markov state $|0\rangle$ leads effectively to an average tunnel-rate from $|0\rangle$
to $|2\rangle$. The same holds true for tunneling in the other direction and tunneling between from $|0\rangle$ to $|1\rangle$.


Next, we consider transitions between model states, which we interpret as tunneling events between quantum states. In general, tunneling is a bidirectional process described by two tunneling rates that must be consistent with the state occupation probabilities in thermodynamic equilibrium. This imposes constraints on model selection. Since Model 2 and 3 are uni-directional for transitions between $|1\rangle$ and $|2\rangle$ and between $|2\rangle$ and $|0\rangle$, they must be excluded. Consequently, only Models 1 and 4 are left which are bidirectional between $|1\rangle$ and $|2\rangle$  and $|2\rangle$ and $|0\rangle$ , respectively. They differ in the pathway through which the excited $(2,0)$ state ($|2\rangle$) can be accessed. In Model 1, direct excitation to the $(2,0)$ state is allowed, whereas in Model 4, excitation occurs exclusively during the tunneling process from the right to the left QD. 

We compare our findings with the analyses of similar systems conducted by Maisi {\it et al.} who explored the dynamics of a double QD system monitoring their charge states using a QPC \cite{MaisiPRL2016}. Like in our study, the QPC measurement exhibited time intervals of high switching rates and intervals with no switching between two current levels (see Fig. 2(f) in \cite{MaisiPRL2016}). In contrast to our data, the system resided mainly in the $(1,1)$ configuration. Moreover, the WTD of their $(1,1)$ configuration exhibited a biexponential decay, whereas the $(2,0)$ configuration showed a monoexponential decay. Three important differences in the experiment may account for the discrepancies between our findings and those of Maisi {\it et al.}:

\textit{Tuning of the QD Levels}: The predominant occupation in the $(1,1)$ charge configuration can be explained by the fact that in Maisi’s experiment, the gate voltage settings made the $(1,1)$ configuration the lowest energy state, whereas in our case, the singlet $(2,0)$ state is the ground state.

\textit{Temperature Effects}: In our experiments, we attribute the occurrence of spin configurations with different energies of the $(2,0)$ configuration to an energy spacing that is comparable with the thermal energy. In contrast, Maisi {\it et al.} conducted their measurements at significantly lower temperatures corresponding to about \SI{5}{\micro eV}, allowing the $(2,0)$ state to behave as a single state due to negligible thermal excitation.

\textit{Magnetic Field Influence}: The data presented by Maisi \textit{et al.}, specifically in their Fig. 2(f), were recorded in the presence of an external magnetic field which increases the energy required for spin flips, making them less likely to occur. Consequently, two distinct timescales for tunneling emerge: tunneling towards the anti-aligned $(2,0)$ state occurs much faster than from the aligned $(1,1)$ state, which requires a spin flip.

Maisi {\it et al.} give a physical motivation for their model where the $(2,0)$ configuration was considered a single state and the $(1,1)$ configuration was split into the four different combinations of spin orientations, $(\uparrow, \uparrow)$, $(\downarrow, \downarrow)$, $(\downarrow, \uparrow)$, and $(\uparrow, \downarrow)$, with a total of eight transitions to account for the complex switching dynamics. We argue that the five-state model of Maisi {\it et al.}, while being physically motivated, cannot be ultimately confirmed by their measurement. Again, the two WTDs only pinpoint four parameters and do not allow for the differentiation of eight transition rates. We found that their measurements can identically be described by only three Markov states $|\Tilde{0}\rangle=(2,0)$, $|\Tilde{1}\rangle=(1,1)_{(\uparrow, \uparrow)/(\downarrow, \downarrow)}$, and $|\Tilde{2}\rangle=(1,1)_{(\uparrow, \downarrow)/(\downarrow, \uparrow)}$ which are connected with modified transition rates $\Tilde{\gamma}_{0,1} = 2\Gamma_\textrm{SO}$, $\Tilde{\gamma}_{1,0} = \Gamma_\textrm{SO}$, $\Tilde{\gamma}_{0,2} = 2\Gamma_\textrm{C}$, $\Tilde{\gamma}_{2,0} = \Gamma_\textrm{C}$, and $\Tilde{\gamma}_{2,1} = \Tilde{\gamma}_{1,2} = 2\Gamma_\textrm{SI}$.

\section{Conclusion}
In this work, we introduced and validated a scheme for the comprehensive characterization of quantum dot arrays using quantum polyspectra and the Akaike information criterion. Our method offers a significant advancement over traditional techniques, such as waiting-time distributions, by relying on the raw detector output of, e.g., a quantum point contact current measurement, instead of quantum jump times. This renders the challenging task of identifying quantum jumps on a background noise unnecessary. Our framework covers the full range between weak and strong measurements and allows for the treatment of coherent and incoherent quantum dynamics. We demonstrated that in the strong measurement regime, quantum polyspectra are on par with a waiting-time analysis in determining transition rates within a double QD system.

Furthermore, we showed that the Akaike Information Criterion can be used to judge models regarding their number of parameters in relation to their power of fitting the measurement traces. Our analysis uncovers hidden excited triplet states in the $(2,0)$ charge configuration. This approach is particularly valuable for exploring larger QD arrays and for investigating unknown environments where the system dynamics are not well-understood. We envision that the application of quantum polyspectra will extend beyond the systems studied here, offering a powerful tool for the analysis and optimization of quantum devices in a wide range of quantum technology applications.

\section*{Acknowledgments}
We acknowledge financial support by the German Science Foundation (DFG) under Project No. 341960391 and No. 510607185 (D.H.), under Germany's Excellence Strategy — EXC 2123 QuantumFrontiers — 390837967 (J.B., R.H.), and the State of Lower Saxony of Germany via the Hannover School for Nanotechnology (J.B.) as well as by the Mercator Research Center Ruhr (MERCUR) under Project No. Ko-2022-0013 (M.S., D.H.).

\appendix

\section{Fourth-order quantum polyspectrum}
\label{app:QuantumPolyspectra}
The fourth-order polyspectrum of the detector output $z(t)$ of the continuously monitored quantum system in the steady state follows from the SME
and the definition of Brillinger's polyspectra $S_z^{(n)}$. 
The fourth-order spectrum (second- and third-order spectra see main text) \cite{footnote1}
\begin{widetext}
	\begin{eqnarray}	
		S_{I_\textrm{QPC}}^{\rm (4)}(\omega_1,\omega_2,\omega_3,\omega_4 = -\omega_1-\omega_2-\omega_3) & = & \hspace{4mm}\beta^8 \hspace{-8mm} \sum_{\{k,l,m,n\} \in \text{prm.} \{1,2,3,4\}}
		\hspace{-8mm} \left[ {\rm Tr}[{\cal A}'{\cal G}'(\omega_n){\cal A}' {\cal G}'(\omega_m + \omega_n){\cal A}'{\cal G}'(\omega_l + \omega_m + \omega_n){\cal A}'\rho_0] \right. \\ \nonumber
		&-& \frac{1}{2 \pi}\int{\rm Tr}[{\cal A}'{\cal G}'(\omega_n) {\cal G}'(\omega_m + \omega_n - \omega){\cal A}'\rho_0]{\rm Tr}[{\cal A}'{\cal G}'(\omega) {\cal G}'(\omega_l +\omega_m + \omega_n){\cal A}'\rho_0]\textrm{d}\omega \\ \nonumber
		&-& \left. \frac{1}{2 \pi}\int{\rm Tr}[{\cal A}'{\cal G}'(\omega_n) {\cal G}'(\omega_l +\omega_m + \omega_n){\cal G}'(\omega_m + \omega_n - \omega){\cal A}'\rho_0]{\rm Tr}[{\cal A}'{\cal G}'(\omega) {\cal A}'\rho_0]\textrm{d}\omega\right]. \label{eq:S4}
	\end{eqnarray}
	was first derived in Refs. \cite{hagelePRB2018,hagelePRB2020E} where also an efficient method for its evaluation was given. 
	These numerical methods are implemented in our QuantumCatch library based on the QuTiP and ArrayFire software libraries \cite{sifftQUANTUMCATCH2023,JOHANSSON20131234,Yalamanchili2015}
\end{widetext}

\section{Spectral value error estimation}
\label{app:ErrorEstimation}
All spectra that are estimated from experimental data exhibit deviations from the true spectrum that only slowly reduce with the square root of the measurement time.
A measure of the error of spectral values needs to be regarded in the fitting procedure to reduce the weight of unreliable parts of the spectrum. Here, we calculate the variance of our experimental spectrum in the following way. The data is divided into $m$ parts of the same size for which separate spectra  $S^{(n)}_j(\vec{\omega})$ with $j=0,…,m-1$ are calculated as described in \cite{sifftPRR2021} where $\vec{\omega}$ regards the frequency dependence of $S^{(n)}_j$
The variance is then estimated via 
\begin{eqnarray}
	{\rm Var}(S^{(n)}) & = & \frac{m}{m-1} \left( \frac{1}{m}\sum_{j=0}^{m-1}
	\left( S^{(n)}_j\right) ^2 - \left( \frac{1}{m}\sum_{j=0}^{m-1} S^{(n)}_j\right)^2 \right) \nonumber \\ 
	& = & \frac{m}{m-1} \left( \overline{(S^{(n)}_j)^2} - \overline{ S^{(n)}_j }^2 \right),
\end{eqnarray}
where the overline indicates an average.

\section{Identical WTDs of Different Three-State Models}
\label{app:TheoreticalWTD}
The WTDs $w_{\textrm{low}}(\tau)$ and $w_{\textrm{high}}(\tau)$ found in the experiment for switching from the low/high to the high/low value exhibit the following
simple structure
\begin{eqnarray}
	\label{eq:general_wtd1}
	w_{\textrm{low}}(\tau) &\propto& e^{- \gamma_1 \tau}, \\ \label{eq:general_wtd2}
	w_{\textrm{high}}(\tau) &\propto& e^{- \gamma_2 \tau} + \alpha e^{- \gamma_3 \tau},
\end{eqnarray}
which is fully characterized by the four parameters $\gamma_1$, $\gamma_2$, $\gamma_3$, and $\alpha$. The omitted prefactors in the equation have to guarantee the normalization condition $\int_0^\infty w_{\textrm{low/high}}(\tau) \, d \tau = 1$ and can therefore not be regarded as additional parameters, as they depend completely on the $\gamma_j$s and $\alpha$.
A general three-state Markov model would include six possible transition rates. 
We will show that such a model yields always WTDs of the form above, implying that no unique three-state model exists.


The WTDs of a general three-state model, in which the states $|1 \rangle$ and $|2 \rangle$ result in the same measurement value, can be found from the dynamics of the density matrix $\rho$ under the general Liouvillian $\mathcal{L}$ 
\begin{eqnarray}
	\label{eq:general_L} 
	\Dot{\rho} & = & \mathcal{L} \rho \nonumber \\
	& = & 
	\left(
	\begin{array}{ccc}
		-\gamma_{01}-\gamma_{02} & \gamma_{10} & \gamma_{20} \\
		\gamma_{01} & -\gamma_{10}-\gamma_{12}& \gamma_{21} \\
		\gamma_{02} & \gamma_{12} & -\gamma_{20}'-\gamma_{21} \\
	\end{array}
	\right) \rho,
\end{eqnarray}
where we have adapted the notation of Brandes in the first line of Eq.~(\ref{eq:general_L}), \cite{Brandes2008}. Here, the equation describes the special case of a continuous Markov model where the density matrix keeps a diagonal form. The density matrix can therefore be represented by a vector $\rho = (\rho_{00}, \rho_{11},\rho_{22})$ and the super operator ${\cal L}$ by a matrix (see second line). Brandes gives a recipe to calculate WTDs for Markov models in terms of ${\cal L}$ and jump operators $\mathcal{J}_\textrm{down}$ and $\mathcal{J}_\textrm{up}$ that are in our case associated with the transitions to the $|0 \rangle$ state and states $|1 \rangle$ or $|2 \rangle$, respectively. Those directly follow from $\mathcal{L}$ as
\begin{eqnarray}
	\mathcal{J}_\textrm{down} & = &
	\left(
	\begin{array}{ccc}
		0 & \gamma_{10} & \gamma_{20} \\
		0 & 0 & 0 \\
		0 & 0 & 0 \\
	\end{array}
	\right) \nonumber 
\end{eqnarray}
and
\begin{eqnarray}
	\mathcal{J}_\textrm{up} & = & 
	\left(
	\begin{array}{ccc}
		0 & 0 & 0 \\
		\gamma_{01} & 0 & 0 \\
		\gamma_{02} & 0 & 0 \\
	\end{array}
	\right). \label{eq:general_jumps}
\end{eqnarray}
\begin{widetext}
	The WTDs of the model can then be found analytically with the help of MATHEMATICA \cite{Brandes2008}
	\begin{eqnarray}
		\label{eq:wtd1}
		w_{\textrm{low}}(\tau) &=& \frac{\operatorname{Tr} \mathcal{J}_\textrm{up} e^{\mathcal{L}_0 \tau} \mathcal{J}_\textrm{down} \rho_0 }{\operatorname{Tr} \mathcal{J}_\textrm{down} \rho_0} =\left(\gamma _{01}+\gamma _{02}\right) e^{- \left(\gamma _{01}+\gamma _{02}\right)\tau }
	\end{eqnarray}
	and
	\begin{eqnarray}
		\label{eq:wtd2}
		w_{\textrm{high}}(\tau) &=& \frac{\operatorname{Tr} \mathcal{J}_{\textrm{down}} e^{\mathcal{L}_0 \tau} \mathcal{J}_{\textrm{up}} \rho_0 }{\operatorname{Tr} \mathcal{J}_{\textrm{up}}  \rho_0  }  = \frac{\gamma _{01} \gamma _{10}+\gamma _{02} \gamma _{20}}{\gamma _{01}+\gamma _{02}}\left[\beta e^{  -\frac{1}{2}  \left(\gamma _{10}+\gamma _{12}+\gamma _{20}+\gamma _{21}-\Gamma \right)\tau }   + \left(1-\beta \right) e^{  -\frac{1}{2}  \left(\gamma _{10}+\gamma _{12}+\gamma _{20}+\gamma _{21}+\Gamma \right)\tau }\right],
	\end{eqnarray}
	with 
	\begin{eqnarray}
		\label{eq:gamma}
		\Gamma &=& \sqrt{2 \gamma _{21} \left(-\gamma _{10}+\gamma _{12}+\gamma _{20}\right)+\left(\gamma _{10}+\gamma _{12}-\gamma _{20}\right){}^2+\left(\gamma_{21} \right)^2} \\
		\beta &=& \frac{\gamma _{01} \left(\gamma _{10} \left(-\gamma _{12}+\gamma _{20}+\gamma _{21}+\Gamma \right)-\left(\gamma \right)_{10}^2+2 \gamma _{12} \gamma _{20}\right)+\gamma _{02} \left(\gamma _{20} \left(\gamma _{10}+\gamma _{12}-\gamma _{21}+\Gamma \right)+2 \gamma _{10} \gamma _{21}-\left(\gamma \right)_{20}^2\right)}{2 \Gamma  \left(\gamma _{01} \gamma _{10}+\gamma _{02} \gamma _{20}\right)}
	\end{eqnarray}
	where $\rho_0 $ is the stationary state with $\mathcal{L} \rho_0 = 0$, i.e., the eigenvector of $\mathcal{L}$ to the zero eigenvalue, and
	$ \mathcal{L}_0 = \mathcal{L} - \mathcal{J}_\textrm{up} - \mathcal{J}_\textrm{down}$ describes 
	the dynamics of the system when no jump occurs in the measured signal. This is e.g. the case when the system dynamics shows only transitions between $|1 \rangle$ and $|2 \rangle$ that result in the same signal level. The analytical WTDs $w_{\textrm{low}}(\tau)$ and $w_{\textrm{high}}(\tau)$ show the same structure as the experimental WTDs shown above.
	
	The six different rates $\gamma_{ij}$ of the general model and the four parameters $\gamma_1$, $\gamma_2$, $\gamma_3$, and $\alpha$ of the experimental WTDs (Eqs.~\eqref{eq:general_wtd1} and \eqref{eq:general_wtd2}) show a non-linear dependency. 
	It is immediately clear that there are many different models that can match the experimental results identically. 
	The numerics of the four optimal models already gave a strong hint, that their WTDs are identical (see Table \ref{tab:rates}). Here, we give an analytical proof. 
	
	We express the parameters of $\mathcal{L}^{(1)}$ (the superscript indicates the model number) by variables $a$, $b$, $c$, and $d$ in a non-linear fashion. That way, the analytical expressions for the WTDs can be expressed without square-roots. We find that a parameterization of the form
	\begin{eqnarray}
		\mathcal{J}_\textrm{down}^{(1)} = 
		\left(
		\begin{array}{ccc}
			0 & \frac{c (a-d) (b+d)}{a b} & 0 \\
			0 & 0 & 0 \\
			0 & 0 & 0 \\
		\end{array}
		\right), \textrm{and } 
		\mathcal{J}_\textrm{up}^{(1)} = 
		\left(
		\begin{array}{ccc}
			0 & 0 & 0 \\
			a & 0 & 0 \\
			0 & 0 & 0 \\
		\end{array}
		\right)
	\end{eqnarray}
	\begin{eqnarray}
		\textrm{with } \mathcal{L}^{(1)} = 
		\left(
		\begin{array}{ccc}
			-a & \frac{c (a-d) (b+d)}{a b} & 0 \\
			a & c \left(\frac{a}{a-d}-\frac{d}{b}-2\right)& \frac{a c}{a-d} \\
			0 & \frac{c d^2 (a-b-d)}{a b (a-d)} & -\frac{a c}{a-d} \\
		\end{array}
		\right);
	\end{eqnarray}
	results in the relatively simple WTDs
	\begin{eqnarray}
		w_{\textrm{low}}(\tau) &= a e^{-a \tau}, \label{eq:proofWTD1}
	\end{eqnarray}
	and
	\begin{eqnarray}
		w_{\textrm{high}}(\tau) &=& \frac{c (b+d)}{a} e^{-c \tau} 
		+ \frac{c (b+d) (a-b-d) }{a b} e^{-\frac{c (b+d) \tau}{b}}. \label{eq:proofWTD2}
	\end{eqnarray}
	One can also find parametrizations for Models 2, 3, and 4 that result in the exact same WTDs. These parametrizations followed from guessing that was guided by the values in Table \ref{tab:rates}. The eleven distinct rates identified across Models 1 to 4 can be reconstructed from merely four base values, illustrating the models’ underlying parameter interdependencies, such as $\SI{1020}{Hz} - \SI{935}{Hz} = \SI{67}{Hz}$, and further $\SI{67}{Hz} \cdot \SI{173}{Hz} / \SI{953}{Hz} \approx \SI{67}{Hz} - \SI{54}{Hz}$. This idea leads to the following parameterizations
	\begin{eqnarray}
		\mathcal{J}_\textrm{down}^{(2)} &=& 
		\left(
		\begin{array}{ccc}
			0 & \frac{c (a-d) (b+d)}{a b} & 0 \\
			0 & 0 & 0 \\
			0 & 0 & 0 \\
		\end{array}
		\right) \quad \text{and} \quad
		\mathcal{J}_\textrm{up}^{(2)} = 
		\left(
		\begin{array}{ccc}
			0 & 0 & 0 \\
			a & 0 & 0 \\
			0 & 0 & 0 \\
		\end{array}
		\right)
		\quad\text{with}\quad
		\mathcal{L}^{(2)} = 
		\left(
		\begin{array}{ccc}
			-a & \frac{c (a-d) (b+d)}{a b} & c \\
			a & -\frac{c (b+d)}{b} & 0 \\
			0 & \frac{c d (b+d)}{a b} & -c \\
		\end{array}
		\right), \nonumber \\
		\mathcal{J}_\textrm{down}^{(3)} &=& 
		\left(
		\begin{array}{ccc}
			0 & \frac{c (b+d)}{b} & 0 \\
			0 & 0 & 0 \\
			0 & 0 & 0 \\
		\end{array}
		\right) \quad\text{and} \quad
		\mathcal{J}_\textrm{up}^{(3)} = 
		\left(
		\begin{array}{ccc}
			0 & 0 & 0 \\
			a-d & 0 & 0 \\
			d & 0 & 0 \\
		\end{array}
		\right)
		\quad \text{with} \quad
		\mathcal{L}^{(3)} = 
		\left(
		\begin{array}{ccc}
			-a & \frac{c (b+d)}{b} & 0 \\
			a-d & -\frac{c (b+d)}{b} & c \\
			d & 0 & -c \\
		\end{array}
		\right),
		\quad\text{and}\quad \nonumber \\
		\mathcal{J}_\textrm{down}^{(4)} &=& 
		\left(
		\begin{array}{ccc}
			0 & \frac{c (b+d)}{b} & c \\
			0 & 0 & 0 \\
			0 & 0 & 0 \\
		\end{array}
		\right)
		\quad\text{and} \quad
		\mathcal{J}_\textrm{up}^{(4)} = 
		\left(
		\begin{array}{ccc}
			0 & 0 & 0 \\
			a-b-d & 0 & 0 \\
			b+d & 0 & 0 \\
		\end{array}
		\right)
		\quad\text{with}\quad
		\mathcal{L}^{(4)} = 
		\left(
		\begin{array}{ccc}
			-a & \frac{c (b+d)}{b} & c \\
			a-b-d & -\frac{c (b+d)}{b} & 0 \\
			b+d & 0 & -c \\
		\end{array}
		\right), \nonumber
	\end{eqnarray}
	that yield the identical WTDs as given by Eq.~(\ref{eq:proofWTD1}) and Eq.~(\ref{eq:proofWTD2}).
	The values in Table \ref{tab:rates} are found for approximately $a \approx \SI{5115}{Hz}$, $b \approx \SI{68}{Hz}$, $c \approx \SI{173}{Hz}$, and $d \approx \SI{332}{Hz}$.
	This implies that given any one of these models, a transformation can be applied to the transition rates that yields a different model with the same WTDs.

	\section{Proof that higher-order WTDs do not provide additional information for a three-state system with two measurement outputs}
	\label{app:ProofHigherOrder}
	Random telegraph noise that is generated by a continuous Markov model with $N$ states and $N$ different output levels 
	can obviously be characterized by $N \times (N-1)$ first-order WTDs that depend on time $\tau$ . 
	In our case of $N = 3$ Markov states, we can distinguish only two detector output levels $I_{\text{high}}$ and $I_{\text{low}}$. Consequently, the three-state model given by Eq.~(\ref{eq:general_L}) and Eq.~(\ref{eq:general_jumps}) can distinguish only two WTDs, Eq.~(\ref{eq:proofWTD1}) and Eq.~(\ref{eq:proofWTD2}). One may therefore ask whether a full characterization of the telegraph noise requires the measurement of second-order or even higher-order waiting-time distributions. Here, we prove that second-order WTDs can in our case of the general three-state model with only two output-levels [Eq.~(\ref{eq:general_L}) and Eq.~(\ref{eq:general_jumps})] always be expressed in terms of their first-order WTDs. Below, we generalize the proof to WTDs of any order.

	
	The two-time WTD for a jump from the low output value to the high output value in time $\tau_1$ and back in time $\tau_2$ is given by
	\begin{equation}
		w_{\text{high, low}}(\tau_2, \tau_1) = \frac{\operatorname{Tr} \left[ \mathcal{J}_{\text{down}} e^{\mathcal{L}_0 \tau_2} \mathcal{J}_{\text{up}} e^{\mathcal{L}_0 \tau_1} \mathcal{J}_{\text{down}} \rho_0 \right]}{\operatorname{Tr} \left[ \mathcal{J}_{\text{down}} \rho_0 \right]}, 
	\end{equation}
	where \( \rho_0 \) is the steady state of the system, \( \mathcal{L}_0 \) is the Liouvillian without jump contributions, and \( \mathcal{J}_{\text{up}} \) and \( \mathcal{J}_{\text{down}} \) are the jump superoperators as defined above.
	The numerator in the equation above can be simplified via the following identity 
	\begin{align}
		\mathcal{J}_{\text{up}} e^{\mathcal{L}_0 \tau_1} \mathcal{J}_{\text{down}} \rho_0 = 
		\left(\gamma _{01}+\gamma _{02}\right) e^{-\tau_0\left(\gamma _{01}+\gamma _{02}\right)} \begin{pmatrix} 
			0 \\
			\frac{\gamma _{01} \left(\gamma _{12} \gamma _{20}+\gamma _{10} \left(\gamma _{20}+\gamma _{21}\right)\right) }{\left(\gamma _{10}+\gamma _{12}\right) \left(\gamma _{02}+\gamma _{20}\right)+\left(\gamma _{02}+\gamma _{10}\right) \gamma _{21}+\gamma _{01} \left(\gamma _{12}+\gamma _{20}+\gamma _{21}\right)} \\
			\frac{\gamma _{02} \left(\gamma _{12} \gamma _{20}+\gamma _{10} \left(\gamma _{20}+\gamma _{21}\right)\right) }{\left(\gamma _{10}+\gamma _{12}\right) \left(\gamma _{02}+\gamma _{20}\right)+\left(\gamma _{02}+\gamma _{10}\right) \gamma _{21}+\gamma _{01} \left(\gamma _{12}+\gamma _{20}+\gamma _{21}\right)}
		\end{pmatrix}
		= w_\text{low}(\tau_1) \mathcal{J}_{\text{up}} \rho_0,
	\end{align}
	which we found via an explicit calculation for the general three-state Markov model using computer algebra, where $w_\text{low}(\tau_1)$ is given by Eq.~(\ref{eq:wtd1}).
	We emphasize that such an identity does generally {\it not} hold for, e.g., a four-state Markov model with two output-levels.	
	Using the identity, we proceed to arrive at 
	\begin{eqnarray}
		w_{\text{high, low}}(\tau_2, \tau_1) & = & \frac{\operatorname{Tr} \left[ \mathcal{J}_{\text{down}} e^{\mathcal{L}_0 \tau_2} \mathcal{J}_{\text{up}} \rho_0 \right] w_\text{low}(\tau_1)}{\operatorname{Tr} \left[ \mathcal{J}_{\text{down}} \rho_0 \right]} \\
		& = & 	\frac{\operatorname{Tr} \left[ \mathcal{J}_{\text{down}} e^{\mathcal{L}_0 \tau_2} \mathcal{J}_{\text{up}} \rho_0 \right] w_\text{low}(\tau_1)}{\operatorname{Tr} \left[ \mathcal{J}_{\text{up}} \rho_0 \right]} \\
		& = & w_\text{high}(\tau_2) w_\text{low}(\tau_1).
	\end{eqnarray}
	In the second line we used another identity, $\operatorname{Tr} \left[ \mathcal{J}_{\text{up}} \rho_0 \right] = \operatorname{Tr} \left[ \mathcal{J}_{\text{down}} \rho_0 \right]$, which we found by explicit calculation for our three-state model. The last line shows that in our case the two-time WTD can always be expressed as a simple product of single-time WTDs.
	
	A similar calulcation yields
	\begin{equation}
		w_{\text{low, high}}(\tau_2, \tau_1) = w_\text{low}(\tau_2) w_\text{high}(\tau_1),
	\end{equation}
	which concludes our proof for the two-time WTDs.

	Also higher-order WTDs turn out to be products of single-time WTDs for three-state Markov models with two measurement outputs. This is shown beginning with the general multi-time WTD
	\begin{equation}
		w_{i_{n}, i_{n-1}, \ldots, i_1}(\tau_{n}, \tau_{n-1}, \ldots, \tau_1) = \frac{\operatorname{Tr} \left[ \mathcal{J}_{i_{n}} e^{\mathcal{L}_0 \tau_{n}} \mathcal{J}_{i_{n-1}} e^{\mathcal{L}_0 \tau_{n-1}} \cdots \mathcal{J}_{i_1} \rho_0 \right]}{\operatorname{Tr} \left[ \mathcal{J}_{i_1} \rho_0 \right]},
	\end{equation}
	with the indices $i_k$ of $w$ and $\mathcal{J}$ taking the values "low" or "high" and "down" or "up", respectively.
	With the identities (see above)
	\begin{eqnarray}
		\mathcal{J}_{\text{up}} e^{\mathcal{L}_0 \tau} \mathcal{J}_{\text{down}} \rho_0 & = & w_\text{low}(\tau) \mathcal{J}_{\text{up}} \rho_0 \nonumber \\
		\mathcal{J}_{\text{down}} e^{\mathcal{L}_0 \tau} \mathcal{J}_{\text{up}} \rho_0 & = & w_\text{high}(\tau) \mathcal{J}_{\text{down}} \rho_0 \nonumber \\
		\operatorname{Tr} \left[ \mathcal{J}_{\text{up}} \rho_0 \right] & = & \operatorname{Tr} \left[ \mathcal{J}_{\text{down}} \rho_0 \right]
	\end{eqnarray}
	we can eliminate the $\mathcal{J}_{i_1}$ operator and find
	\begin{equation}
		w_{i_{n}, i_{n-1}, \ldots, i_1}(\tau_{n}, \tau_{n-1}, \ldots, \tau_1) = \frac{\operatorname{Tr} \left[ \mathcal{J}_{i_{n}} e^{\mathcal{L}_0 \tau_{n}} \mathcal{J}_{i_{n-1}} e^{\mathcal{L}_0 \tau_{n-1}} \cdots \mathcal{J}_{i_2} \rho_0 \right]}{\operatorname{Tr} \left[ \mathcal{J}_{i_2} \rho_0 \right]}
		w_{i_1}(\tau_1).
	\end{equation}
	Repeating the corresponding procedure for $i_2$, $i_3$,..., $i_n$, we find
	\begin{equation}
		w_{i_{n}, i_{n-1}, \ldots, i_1}(\tau_{n}, \tau_{n-1}, \ldots, \tau_1) = \prod_{k=1}^{n} w_{i_k}(\tau_k).
	\end{equation}	
	This means, that any random telegraph noise of a three-state Markov system is fully characterized by its single-time WTDs. The higher-order WTDs do not contain additional information about the statistics of telegraph noise. Any two such systems with identical waiting-time distributions and same signal levels will also exhibit identical polyspectra. This statement, however, does not generalize to larger Markov models or periodically driven systems \cite{BayerArxiv2024}.
\end{widetext}


\begin{thebibliography}{59}%
	\makeatletter
	\providecommand \@ifxundefined [1]{%
		\@ifx{#1\undefined}
	}%
	\providecommand \@ifnum [1]{%
		\ifnum #1\expandafter \@firstoftwo
		\else \expandafter \@secondoftwo
		\fi
	}%
	\providecommand \@ifx [1]{%
		\ifx #1\expandafter \@firstoftwo
		\else \expandafter \@secondoftwo
		\fi
	}%
	\providecommand \natexlab [1]{#1}%
	\providecommand \enquote  [1]{``#1''}%
	\providecommand \bibnamefont  [1]{#1}%
	\providecommand \bibfnamefont [1]{#1}%
	\providecommand \citenamefont [1]{#1}%
	\providecommand \href@noop [0]{\@secondoftwo}%
	\providecommand \href [0]{\begingroup \@sanitize@url \@href}%
	\providecommand \@href[1]{\@@startlink{#1}\@@href}%
	\providecommand \@@href[1]{\endgroup#1\@@endlink}%
	\providecommand \@sanitize@url [0]{\catcode `\\12\catcode `\$12\catcode
		`\&12\catcode `\#12\catcode `\^12\catcode `\_12\catcode `\%12\relax}%
	\providecommand \@@startlink[1]{}%
	\providecommand \@@endlink[0]{}%
	\providecommand \url  [0]{\begingroup\@sanitize@url \@url }%
	\providecommand \@url [1]{\endgroup\@href {#1}{\urlprefix }}%
	\providecommand \urlprefix  [0]{URL }%
	\providecommand \Eprint [0]{\href }%
	\providecommand \doibase [0]{https://doi.org/}%
	\providecommand \selectlanguage [0]{\@gobble}%
	\providecommand \bibinfo  [0]{\@secondoftwo}%
	\providecommand \bibfield  [0]{\@secondoftwo}%
	\providecommand \translation [1]{[#1]}%
	\providecommand \BibitemOpen [0]{}%
	\providecommand \bibitemStop [0]{}%
	\providecommand \bibitemNoStop [0]{.\EOS\space}%
	\providecommand \EOS [0]{\spacefactor3000\relax}%
	\providecommand \BibitemShut  [1]{\csname bibitem#1\endcsname}%
	\let\auto@bib@innerbib\@empty
	\bibitem [{\citenamefont {Heindel}\ \emph {et~al.}(2012)\citenamefont
		{Heindel}, \citenamefont {Kessler}, \citenamefont {Rau}, \citenamefont
		{Schneider}, \citenamefont {Fürst}, \citenamefont {Hargart}, \citenamefont
		{Schulz}, \citenamefont {Eichfelder}, \citenamefont {Roßbach}, \citenamefont
		{Nauerth}, \citenamefont {Lermer}, \citenamefont {Weier}, \citenamefont
		{Jetter}, \citenamefont {Kamp}, \citenamefont {Reitzenstein}, \citenamefont
		{Höfling}, \citenamefont {Michler}, \citenamefont {Weinfurter},\ and\
		\citenamefont {Forchel}}]{HeindelNJP2012}%
	\BibitemOpen
	\bibfield  {author} {\bibinfo {author} {\bibfnamefont {T.}~\bibnamefont
			{Heindel}}, \bibinfo {author} {\bibfnamefont {C.~A.}\ \bibnamefont
			{Kessler}}, \bibinfo {author} {\bibfnamefont {M.}~\bibnamefont {Rau}},
		\bibinfo {author} {\bibfnamefont {C.}~\bibnamefont {Schneider}}, \bibinfo
		{author} {\bibfnamefont {M.}~\bibnamefont {Fürst}}, \bibinfo {author}
		{\bibfnamefont {F.}~\bibnamefont {Hargart}}, \bibinfo {author} {\bibfnamefont
			{W.-M.}\ \bibnamefont {Schulz}}, \bibinfo {author} {\bibfnamefont
			{M.}~\bibnamefont {Eichfelder}}, \bibinfo {author} {\bibfnamefont
			{R.}~\bibnamefont {Roßbach}}, \bibinfo {author} {\bibfnamefont
			{S.}~\bibnamefont {Nauerth}}, \bibinfo {author} {\bibfnamefont
			{M.}~\bibnamefont {Lermer}}, \bibinfo {author} {\bibfnamefont
			{H.}~\bibnamefont {Weier}}, \bibinfo {author} {\bibfnamefont
			{M.}~\bibnamefont {Jetter}}, \bibinfo {author} {\bibfnamefont
			{M.}~\bibnamefont {Kamp}}, \bibinfo {author} {\bibfnamefont {S.}~\bibnamefont
			{Reitzenstein}}, \bibinfo {author} {\bibfnamefont {S.}~\bibnamefont
			{Höfling}}, \bibinfo {author} {\bibfnamefont {P.}~\bibnamefont {Michler}},
		\bibinfo {author} {\bibfnamefont {H.}~\bibnamefont {Weinfurter}},\ and\
		\bibinfo {author} {\bibfnamefont {A.}~\bibnamefont {Forchel}},\ }\bibfield
	{title} {\bibinfo {title} {Quantum key distribution using quantum dot
			single-photon emitting diodes in the red and near infrared spectral range},\
	}\href {https://doi.org/10.1088/1367-2630/14/8/083001} {\bibfield  {journal}
		{\bibinfo  {journal} {New Journal of Physics}\ }\textbf {\bibinfo {volume}
			{14}},\ \bibinfo {pages} {083001} (\bibinfo {year} {2012})}\BibitemShut
	{NoStop}%
	\bibitem [{\citenamefont {M{\"u}ller}\ \emph {et~al.}(2014)\citenamefont
		{M{\"u}ller}, \citenamefont {Bounouar}, \citenamefont {J{\"o}ns},
		\citenamefont {Gl{\"a}ssl},\ and\ \citenamefont
		{Michler}}]{MuellerNatPho2014}%
	\BibitemOpen
	\bibfield  {author} {\bibinfo {author} {\bibfnamefont {M.}~\bibnamefont
			{M{\"u}ller}}, \bibinfo {author} {\bibfnamefont {S.}~\bibnamefont
			{Bounouar}}, \bibinfo {author} {\bibfnamefont {K.~D.}\ \bibnamefont
			{J{\"o}ns}}, \bibinfo {author} {\bibfnamefont {M.}~\bibnamefont
			{Gl{\"a}ssl}},\ and\ \bibinfo {author} {\bibfnamefont {P.}~\bibnamefont
			{Michler}},\ }\bibfield  {title} {\bibinfo {title} {On-demand generation of
			indistinguishable polarization-entangled photon pairs},\ }\href
	{https://doi.org/10.1038/nphoton.2013.377} {\bibfield  {journal} {\bibinfo
			{journal} {Nature Photonics}\ }\textbf {\bibinfo {volume} {8}},\ \bibinfo
		{pages} {224} (\bibinfo {year} {2014})}\BibitemShut {NoStop}%
	\bibitem [{\citenamefont {Gschrey}\ \emph {et~al.}(2015)\citenamefont
		{Gschrey}, \citenamefont {Thoma}, \citenamefont {Schnauber}, \citenamefont
		{Seifried}, \citenamefont {Schmidt}, \citenamefont {Wohlfeil}, \citenamefont
		{Kr{\"u}ger}, \citenamefont {Schulze}, \citenamefont {Heindel}, \citenamefont
		{Burger}, \citenamefont {Schmidt}, \citenamefont {Strittmatter},
		\citenamefont {Rodt},\ and\ \citenamefont
		{Reitzenstein}}]{GschreyNatComm2015}%
	\BibitemOpen
	\bibfield  {author} {\bibinfo {author} {\bibfnamefont {M.}~\bibnamefont
			{Gschrey}}, \bibinfo {author} {\bibfnamefont {A.}~\bibnamefont {Thoma}},
		\bibinfo {author} {\bibfnamefont {P.}~\bibnamefont {Schnauber}}, \bibinfo
		{author} {\bibfnamefont {M.}~\bibnamefont {Seifried}}, \bibinfo {author}
		{\bibfnamefont {R.}~\bibnamefont {Schmidt}}, \bibinfo {author} {\bibfnamefont
			{B.}~\bibnamefont {Wohlfeil}}, \bibinfo {author} {\bibfnamefont
			{L.}~\bibnamefont {Kr{\"u}ger}}, \bibinfo {author} {\bibfnamefont {J.~H.}\
			\bibnamefont {Schulze}}, \bibinfo {author} {\bibfnamefont {T.}~\bibnamefont
			{Heindel}}, \bibinfo {author} {\bibfnamefont {S.}~\bibnamefont {Burger}},
		\bibinfo {author} {\bibfnamefont {F.}~\bibnamefont {Schmidt}}, \bibinfo
		{author} {\bibfnamefont {A.}~\bibnamefont {Strittmatter}}, \bibinfo {author}
		{\bibfnamefont {S.}~\bibnamefont {Rodt}},\ and\ \bibinfo {author}
		{\bibfnamefont {S.}~\bibnamefont {Reitzenstein}},\ }\bibfield  {title}
	{\bibinfo {title} {Highly indistinguishable photons from deterministic
			quantum-dot microlenses utilizing three-dimensional in situ electron-beam
			lithography},\ }\href {https://doi.org/10.1038/ncomms8662} {\bibfield
		{journal} {\bibinfo  {journal} {Nature Communications}\ }\textbf {\bibinfo
			{volume} {6}},\ \bibinfo {pages} {7662} (\bibinfo {year} {2015})}\BibitemShut
	{NoStop}%
	\bibitem [{\citenamefont {Senellart}\ \emph {et~al.}(2017)\citenamefont
		{Senellart}, \citenamefont {Solomon},\ and\ \citenamefont
		{White}}]{SenellartNNANO2017}%
	\BibitemOpen
	\bibfield  {author} {\bibinfo {author} {\bibfnamefont {P.}~\bibnamefont
			{Senellart}}, \bibinfo {author} {\bibfnamefont {G.}~\bibnamefont {Solomon}},\
		and\ \bibinfo {author} {\bibfnamefont {A.}~\bibnamefont {White}},\ }\bibfield
	{title} {\bibinfo {title} {High-performance semiconductor quantum-dot
			single-photon sources},\ }\href {https://doi.org/10.1038/nnano.2017.218}
	{\bibfield  {journal} {\bibinfo  {journal} {Nature Nanotechnology}\ }\textbf
		{\bibinfo {volume} {12}},\ \bibinfo {pages} {1026} (\bibinfo {year}
		{2017})}\BibitemShut {NoStop}%
	\bibitem [{\citenamefont {Tomm}\ \emph {et~al.}(2021)\citenamefont {Tomm},
		\citenamefont {Javadi}, \citenamefont {Antoniadis}, \citenamefont {Najer},
		\citenamefont {L{\"o}bl}, \citenamefont {Korsch}, \citenamefont {Schott},
		\citenamefont {Valentin}, \citenamefont {Wieck}, \citenamefont {Ludwig},\
		and\ \citenamefont {Warburton}}]{TommNatNano2021}%
	\BibitemOpen
	\bibfield  {author} {\bibinfo {author} {\bibfnamefont {N.}~\bibnamefont
			{Tomm}}, \bibinfo {author} {\bibfnamefont {A.}~\bibnamefont {Javadi}},
		\bibinfo {author} {\bibfnamefont {N.~O.}\ \bibnamefont {Antoniadis}},
		\bibinfo {author} {\bibfnamefont {D.}~\bibnamefont {Najer}}, \bibinfo
		{author} {\bibfnamefont {M.~C.}\ \bibnamefont {L{\"o}bl}}, \bibinfo {author}
		{\bibfnamefont {A.~R.}\ \bibnamefont {Korsch}}, \bibinfo {author}
		{\bibfnamefont {R.}~\bibnamefont {Schott}}, \bibinfo {author} {\bibfnamefont
			{S.~R.}\ \bibnamefont {Valentin}}, \bibinfo {author} {\bibfnamefont {A.~D.}\
			\bibnamefont {Wieck}}, \bibinfo {author} {\bibfnamefont {A.}~\bibnamefont
			{Ludwig}},\ and\ \bibinfo {author} {\bibfnamefont {R.~J.}\ \bibnamefont
			{Warburton}},\ }\bibfield  {title} {\bibinfo {title} {A bright and fast
			source of coherent single photons},\ }\href
	{https://doi.org/10.1038/s41565-020-00831-x} {\bibfield  {journal} {\bibinfo
			{journal} {Nature Nanotechnology}\ }\textbf {\bibinfo {volume} {16}},\
		\bibinfo {pages} {399} (\bibinfo {year} {2021})}\BibitemShut {NoStop}%
	\bibitem [{\citenamefont {Zhai}\ \emph {et~al.}(2022)\citenamefont {Zhai},
		\citenamefont {Nguyen}, \citenamefont {Spinnler}, \citenamefont {Ritzmann},
		\citenamefont {L{\"o}bl}, \citenamefont {Wieck}, \citenamefont {Ludwig},
		\citenamefont {Javadi},\ and\ \citenamefont {Warburton}}]{ZhaiNatNano2022}%
	\BibitemOpen
	\bibfield  {author} {\bibinfo {author} {\bibfnamefont {L.}~\bibnamefont
			{Zhai}}, \bibinfo {author} {\bibfnamefont {G.~N.}\ \bibnamefont {Nguyen}},
		\bibinfo {author} {\bibfnamefont {C.}~\bibnamefont {Spinnler}}, \bibinfo
		{author} {\bibfnamefont {J.}~\bibnamefont {Ritzmann}}, \bibinfo {author}
		{\bibfnamefont {M.~C.}\ \bibnamefont {L{\"o}bl}}, \bibinfo {author}
		{\bibfnamefont {A.~D.}\ \bibnamefont {Wieck}}, \bibinfo {author}
		{\bibfnamefont {A.}~\bibnamefont {Ludwig}}, \bibinfo {author} {\bibfnamefont
			{A.}~\bibnamefont {Javadi}},\ and\ \bibinfo {author} {\bibfnamefont {R.~J.}\
			\bibnamefont {Warburton}},\ }\bibfield  {title} {\bibinfo {title} {Quantum
			interference of identical photons from remote gaas quantum dots},\ }\href
	{https://doi.org/10.1038/s41565-022-01131-2} {\bibfield  {journal} {\bibinfo
			{journal} {Nature Nanotechnology}\ }\textbf {\bibinfo {volume} {17}},\
		\bibinfo {pages} {829} (\bibinfo {year} {2022})}\BibitemShut {NoStop}%
	\bibitem [{\citenamefont {Zhang}\ \emph {et~al.}(2023)\citenamefont {Zhang},
		\citenamefont {Cai},\ and\ \citenamefont {Chan}}]{ZhangBiosensors2023}%
	\BibitemOpen
	\bibfield  {author} {\bibinfo {author} {\bibfnamefont {Y.}~\bibnamefont
			{Zhang}}, \bibinfo {author} {\bibfnamefont {N.}~\bibnamefont {Cai}},\ and\
		\bibinfo {author} {\bibfnamefont {V.}~\bibnamefont {Chan}},\ }\bibfield
	{title} {\bibinfo {title} {Recent advances in silicon quantum dot-based
			fluorescent biosensors},\ }\bibfield  {journal} {\bibinfo  {journal}
		{Biosensors}\ }\textbf {\bibinfo {volume} {13}},\ \href
	{https://doi.org/10.3390/bios13030311} {10.3390/bios13030311} (\bibinfo
	{year} {2023})\BibitemShut {NoStop}%
	\bibitem [{\citenamefont {Boeyens}\ \emph {et~al.}(2023)\citenamefont
		{Boeyens}, \citenamefont {Annby-Andersson}, \citenamefont {Bakhshinezhad},
		\citenamefont {Haack}, \citenamefont {Perarnau-Llobet}, \citenamefont
		{Nimmrichter}, \citenamefont {Potts},\ and\ \citenamefont
		{Mehboudi}}]{Boeyens_2023}%
	\BibitemOpen
	\bibfield  {author} {\bibinfo {author} {\bibfnamefont {J.}~\bibnamefont
			{Boeyens}}, \bibinfo {author} {\bibfnamefont {B.}~\bibnamefont
			{Annby-Andersson}}, \bibinfo {author} {\bibfnamefont {P.}~\bibnamefont
			{Bakhshinezhad}}, \bibinfo {author} {\bibfnamefont {G.}~\bibnamefont
			{Haack}}, \bibinfo {author} {\bibfnamefont {M.}~\bibnamefont
			{Perarnau-Llobet}}, \bibinfo {author} {\bibfnamefont {S.}~\bibnamefont
			{Nimmrichter}}, \bibinfo {author} {\bibfnamefont {P.~P.}\ \bibnamefont
			{Potts}},\ and\ \bibinfo {author} {\bibfnamefont {M.}~\bibnamefont
			{Mehboudi}},\ }\bibfield  {title} {\bibinfo {title} {Probe thermometry with
			continuous measurements},\ }\href {https://doi.org/10.1088/1367-2630/ad0e8a}
	{\bibfield  {journal} {\bibinfo  {journal} {New Journal of Physics}\ }\textbf
		{\bibinfo {volume} {25}},\ \bibinfo {pages} {123009} (\bibinfo {year}
		{2023})}\BibitemShut {NoStop}%
	\bibitem [{\citenamefont {Loss}\ and\ \citenamefont
		{DiVincenzo}(1998)}]{LossPRA1998}%
	\BibitemOpen
	\bibfield  {author} {\bibinfo {author} {\bibfnamefont {D.}~\bibnamefont
			{Loss}}\ and\ \bibinfo {author} {\bibfnamefont {D.~P.}\ \bibnamefont
			{DiVincenzo}},\ }\bibfield  {title} {\bibinfo {title} {Quantum computation
			with quantum dots},\ }\href {https://doi.org/10.1103/PhysRevA.57.120}
	{\bibfield  {journal} {\bibinfo  {journal} {Phys. Rev. A}\ }\textbf {\bibinfo
			{volume} {57}},\ \bibinfo {pages} {120} (\bibinfo {year} {1998})}\BibitemShut
	{NoStop}%
	\bibitem [{\citenamefont {Petta}\ \emph {et~al.}(2005)\citenamefont {Petta},
		\citenamefont {Johnson}, \citenamefont {Taylor}, \citenamefont {Laird},
		\citenamefont {Yacoby}, \citenamefont {Lukin}, \citenamefont {Marcus},
		\citenamefont {Hanson},\ and\ \citenamefont {Gossard}}]{PettaScience2005}%
	\BibitemOpen
	\bibfield  {author} {\bibinfo {author} {\bibfnamefont {J.~R.}\ \bibnamefont
			{Petta}}, \bibinfo {author} {\bibfnamefont {A.~C.}\ \bibnamefont {Johnson}},
		\bibinfo {author} {\bibfnamefont {J.~M.}\ \bibnamefont {Taylor}}, \bibinfo
		{author} {\bibfnamefont {E.~A.}\ \bibnamefont {Laird}}, \bibinfo {author}
		{\bibfnamefont {A.}~\bibnamefont {Yacoby}}, \bibinfo {author} {\bibfnamefont
			{M.~D.}\ \bibnamefont {Lukin}}, \bibinfo {author} {\bibfnamefont {C.~M.}\
			\bibnamefont {Marcus}}, \bibinfo {author} {\bibfnamefont {M.~P.}\
			\bibnamefont {Hanson}},\ and\ \bibinfo {author} {\bibfnamefont {A.~C.}\
			\bibnamefont {Gossard}},\ }\bibfield  {title} {\bibinfo {title} {Coherent
			manipulation of coupled electron spins in semiconductor quantum dots},\
	}\href {https://doi.org/10.1126/science.1116955} {\bibfield  {journal}
		{\bibinfo  {journal} {Science}\ }\textbf {\bibinfo {volume} {309}},\ \bibinfo
		{pages} {2180} (\bibinfo {year} {2005})},\ \Eprint
	{https://arxiv.org/abs/https://www.science.org/doi/pdf/10.1126/science.1116955}
	{https://www.science.org/doi/pdf/10.1126/science.1116955} \BibitemShut
	{NoStop}%
	\bibitem [{\citenamefont {Philips}\ \emph {et~al.}(2022)\citenamefont
		{Philips}, \citenamefont {M{\k a}dzik}, \citenamefont {Amitonov},
		\citenamefont {de~Snoo}, \citenamefont {Russ}, \citenamefont {Kalhor},
		\citenamefont {Volk}, \citenamefont {Lawrie}, \citenamefont {Brousse},
		\citenamefont {Tryputen}, \citenamefont {Wuetz}, \citenamefont {Sammak},
		\citenamefont {Veldhorst}, \citenamefont {Scappucci},\ and\ \citenamefont
		{Vandersypen}}]{PhillipsNATURE2022}%
	\BibitemOpen
	\bibfield  {author} {\bibinfo {author} {\bibfnamefont {S.~G.~J.}\
			\bibnamefont {Philips}}, \bibinfo {author} {\bibfnamefont {M.~T.}\
			\bibnamefont {M{\k a}dzik}}, \bibinfo {author} {\bibfnamefont {S.~V.}\
			\bibnamefont {Amitonov}}, \bibinfo {author} {\bibfnamefont {S.~L.}\
			\bibnamefont {de~Snoo}}, \bibinfo {author} {\bibfnamefont {M.}~\bibnamefont
			{Russ}}, \bibinfo {author} {\bibfnamefont {N.}~\bibnamefont {Kalhor}},
		\bibinfo {author} {\bibfnamefont {C.}~\bibnamefont {Volk}}, \bibinfo {author}
		{\bibfnamefont {W.~I.~L.}\ \bibnamefont {Lawrie}}, \bibinfo {author}
		{\bibfnamefont {D.}~\bibnamefont {Brousse}}, \bibinfo {author} {\bibfnamefont
			{L.}~\bibnamefont {Tryputen}}, \bibinfo {author} {\bibfnamefont {B.~P.}\
			\bibnamefont {Wuetz}}, \bibinfo {author} {\bibfnamefont {A.}~\bibnamefont
			{Sammak}}, \bibinfo {author} {\bibfnamefont {M.}~\bibnamefont {Veldhorst}},
		\bibinfo {author} {\bibfnamefont {G.}~\bibnamefont {Scappucci}},\ and\
		\bibinfo {author} {\bibfnamefont {L.~M.~K.}\ \bibnamefont {Vandersypen}},\
	}\bibfield  {title} {\bibinfo {title} {Universal control of a six-qubit
			quantum processor in silicon},\ }\href
	{https://doi.org/10.1038/s41586-022-05117-x} {\bibfield  {journal} {\bibinfo
			{journal} {Nature}\ }\textbf {\bibinfo {volume} {609}},\ \bibinfo {pages}
		{919} (\bibinfo {year} {2022})}\BibitemShut {NoStop}%
	\bibitem [{\citenamefont {Efros}\ and\ \citenamefont
		{Rosen}(1997)}]{EfrosPRL1997}%
	\BibitemOpen
	\bibfield  {author} {\bibinfo {author} {\bibfnamefont {A.~L.}\ \bibnamefont
			{Efros}}\ and\ \bibinfo {author} {\bibfnamefont {M.}~\bibnamefont {Rosen}},\
	}\bibfield  {title} {\bibinfo {title} {Random telegraph signal in the
			photoluminescence intensity of a single quantum dot},\ }\href
	{https://doi.org/10.1103/PhysRevLett.78.1110} {\bibfield  {journal} {\bibinfo
			{journal} {Phys. Rev. Lett.}\ }\textbf {\bibinfo {volume} {78}},\ \bibinfo
		{pages} {1110} (\bibinfo {year} {1997})}\BibitemShut {NoStop}%
	\bibitem [{\citenamefont {Jahn}\ \emph {et~al.}(2015)\citenamefont {Jahn},
		\citenamefont {Munsch}, \citenamefont {B\'eguin}, \citenamefont {Kuhlmann},
		\citenamefont {Renggli}, \citenamefont {Huo}, \citenamefont {Ding},
		\citenamefont {Trotta}, \citenamefont {Reindl}, \citenamefont {Schmidt},
		\citenamefont {Rastelli}, \citenamefont {Treutlein},\ and\ \citenamefont
		{Warburton}}]{JahnPRB2015}%
	\BibitemOpen
	\bibfield  {author} {\bibinfo {author} {\bibfnamefont {J.-P.}\ \bibnamefont
			{Jahn}}, \bibinfo {author} {\bibfnamefont {M.}~\bibnamefont {Munsch}},
		\bibinfo {author} {\bibfnamefont {L.}~\bibnamefont {B\'eguin}}, \bibinfo
		{author} {\bibfnamefont {A.~V.}\ \bibnamefont {Kuhlmann}}, \bibinfo {author}
		{\bibfnamefont {M.}~\bibnamefont {Renggli}}, \bibinfo {author} {\bibfnamefont
			{Y.}~\bibnamefont {Huo}}, \bibinfo {author} {\bibfnamefont {F.}~\bibnamefont
			{Ding}}, \bibinfo {author} {\bibfnamefont {R.}~\bibnamefont {Trotta}},
		\bibinfo {author} {\bibfnamefont {M.}~\bibnamefont {Reindl}}, \bibinfo
		{author} {\bibfnamefont {O.~G.}\ \bibnamefont {Schmidt}}, \bibinfo {author}
		{\bibfnamefont {A.}~\bibnamefont {Rastelli}}, \bibinfo {author}
		{\bibfnamefont {P.}~\bibnamefont {Treutlein}},\ and\ \bibinfo {author}
		{\bibfnamefont {R.~J.}\ \bibnamefont {Warburton}},\ }\bibfield  {title}
	{\bibinfo {title} {An artificial rb atom in a semiconductor with
			lifetime-limited linewidth},\ }\href
	{https://doi.org/10.1103/PhysRevB.92.245439} {\bibfield  {journal} {\bibinfo
			{journal} {Phys. Rev. B}\ }\textbf {\bibinfo {volume} {92}},\ \bibinfo
		{pages} {245439} (\bibinfo {year} {2015})}\BibitemShut {NoStop}%
	\bibitem [{\citenamefont {Santori}\ \emph {et~al.}(2002)\citenamefont
		{Santori}, \citenamefont {Fattal}, \citenamefont {Vu{\v c}kovi{\'c}},
		\citenamefont {Solomon},\ and\ \citenamefont {Yamamoto}}]{SantoriNature2002}%
	\BibitemOpen
	\bibfield  {author} {\bibinfo {author} {\bibfnamefont {C.}~\bibnamefont
			{Santori}}, \bibinfo {author} {\bibfnamefont {D.}~\bibnamefont {Fattal}},
		\bibinfo {author} {\bibfnamefont {J.}~\bibnamefont {Vu{\v c}kovi{\'c}}},
		\bibinfo {author} {\bibfnamefont {G.~S.}\ \bibnamefont {Solomon}},\ and\
		\bibinfo {author} {\bibfnamefont {Y.}~\bibnamefont {Yamamoto}},\ }\bibfield
	{title} {\bibinfo {title} {Indistinguishable photons from a single-photon
			device},\ }\href {https://doi.org/10.1038/nature01086} {\bibfield  {journal}
		{\bibinfo  {journal} {Nature}\ }\textbf {\bibinfo {volume} {419}},\ \bibinfo
		{pages} {594} (\bibinfo {year} {2002})}\BibitemShut {NoStop}%
	\bibitem [{\citenamefont {Warburton}(2013)}]{WarburtonNatMat2013}%
	\BibitemOpen
	\bibfield  {author} {\bibinfo {author} {\bibfnamefont {R.~J.}\ \bibnamefont
			{Warburton}},\ }\bibfield  {title} {\bibinfo {title} {Single spins in
			self-assembled quantum dots},\ }\href {https://doi.org/10.1038/nmat3585}
	{\bibfield  {journal} {\bibinfo  {journal} {Nature Materials}\ }\textbf
		{\bibinfo {volume} {12}},\ \bibinfo {pages} {483} (\bibinfo {year}
		{2013})}\BibitemShut {NoStop}%
	\bibitem [{\citenamefont {Nguyen}\ \emph {et~al.}(2023)\citenamefont {Nguyen},
		\citenamefont {Spinnler}, \citenamefont {Hogg}, \citenamefont {Zhai},
		\citenamefont {Javadi}, \citenamefont {Schrader}, \citenamefont {Erbe},
		\citenamefont {Wyss}, \citenamefont {Ritzmann}, \citenamefont {Babin},
		\citenamefont {Wieck}, \citenamefont {Ludwig},\ and\ \citenamefont
		{Warburton}}]{NguyenPRL2023}%
	\BibitemOpen
	\bibfield  {author} {\bibinfo {author} {\bibfnamefont {G.~N.}\ \bibnamefont
			{Nguyen}}, \bibinfo {author} {\bibfnamefont {C.}~\bibnamefont {Spinnler}},
		\bibinfo {author} {\bibfnamefont {M.~R.}\ \bibnamefont {Hogg}}, \bibinfo
		{author} {\bibfnamefont {L.}~\bibnamefont {Zhai}}, \bibinfo {author}
		{\bibfnamefont {A.}~\bibnamefont {Javadi}}, \bibinfo {author} {\bibfnamefont
			{C.~A.}\ \bibnamefont {Schrader}}, \bibinfo {author} {\bibfnamefont
			{M.}~\bibnamefont {Erbe}}, \bibinfo {author} {\bibfnamefont {M.}~\bibnamefont
			{Wyss}}, \bibinfo {author} {\bibfnamefont {J.}~\bibnamefont {Ritzmann}},
		\bibinfo {author} {\bibfnamefont {H.-G.}\ \bibnamefont {Babin}}, \bibinfo
		{author} {\bibfnamefont {A.~D.}\ \bibnamefont {Wieck}}, \bibinfo {author}
		{\bibfnamefont {A.}~\bibnamefont {Ludwig}},\ and\ \bibinfo {author}
		{\bibfnamefont {R.~J.}\ \bibnamefont {Warburton}},\ }\bibfield  {title}
	{\bibinfo {title} {Enhanced electron-spin coherence in a gaas quantum
			emitter},\ }\href {https://doi.org/10.1103/PhysRevLett.131.210805} {\bibfield
		{journal} {\bibinfo  {journal} {Phys. Rev. Lett.}\ }\textbf {\bibinfo
			{volume} {131}},\ \bibinfo {pages} {210805} (\bibinfo {year}
		{2023})}\BibitemShut {NoStop}%
	\bibitem [{\citenamefont {Hanson}\ \emph {et~al.}(2007)\citenamefont {Hanson},
		\citenamefont {Kouwenhoven}, \citenamefont {Petta}, \citenamefont {Tarucha},\
		and\ \citenamefont {Vandersypen}}]{HansonRevModPhys2007}%
	\BibitemOpen
	\bibfield  {author} {\bibinfo {author} {\bibfnamefont {R.}~\bibnamefont
			{Hanson}}, \bibinfo {author} {\bibfnamefont {L.~P.}\ \bibnamefont
			{Kouwenhoven}}, \bibinfo {author} {\bibfnamefont {J.~R.}\ \bibnamefont
			{Petta}}, \bibinfo {author} {\bibfnamefont {S.}~\bibnamefont {Tarucha}},\
		and\ \bibinfo {author} {\bibfnamefont {L.~M.~K.}\ \bibnamefont
			{Vandersypen}},\ }\bibfield  {title} {\bibinfo {title} {Spins in few-electron
			quantum dots},\ }\href {https://doi.org/10.1103/RevModPhys.79.1217}
	{\bibfield  {journal} {\bibinfo  {journal} {Rev. Mod. Phys.}\ }\textbf
		{\bibinfo {volume} {79}},\ \bibinfo {pages} {1217} (\bibinfo {year}
		{2007})}\BibitemShut {NoStop}%
	\bibitem [{\citenamefont {Nowack}\ \emph {et~al.}(2007)\citenamefont {Nowack},
		\citenamefont {Koppens}, \citenamefont {Nazarov},\ and\ \citenamefont
		{Vandersypen}}]{NowackScience2011}%
	\BibitemOpen
	\bibfield  {author} {\bibinfo {author} {\bibfnamefont {K.~C.}\ \bibnamefont
			{Nowack}}, \bibinfo {author} {\bibfnamefont {F.~H.~L.}\ \bibnamefont
			{Koppens}}, \bibinfo {author} {\bibfnamefont {Y.~V.}\ \bibnamefont
			{Nazarov}},\ and\ \bibinfo {author} {\bibfnamefont {L.~M.~K.}\ \bibnamefont
			{Vandersypen}},\ }\bibfield  {title} {\bibinfo {title} {Coherent control of a
			single electron spin with electric fields},\ }\href
	{https://doi.org/10.1126/science.1148092} {\bibfield  {journal} {\bibinfo
			{journal} {Science}\ }\textbf {\bibinfo {volume} {318}},\ \bibinfo {pages}
		{1430} (\bibinfo {year} {2007})},\ \Eprint
	{https://arxiv.org/abs/https://www.science.org/doi/pdf/10.1126/science.1148092}
	{https://www.science.org/doi/pdf/10.1126/science.1148092} \BibitemShut
	{NoStop}%
	\bibitem [{\citenamefont {Watson}\ \emph {et~al.}(2018)\citenamefont {Watson},
		\citenamefont {Philips}, \citenamefont {Kawakami}, \citenamefont {Ward},
		\citenamefont {Scarlino}, \citenamefont {Veldhorst}, \citenamefont {Savage},
		\citenamefont {Lagally}, \citenamefont {Friesen}, \citenamefont
		{Coppersmith}, \citenamefont {Eriksson},\ and\ \citenamefont
		{Vandersypen}}]{WatsonNature2018}%
	\BibitemOpen
	\bibfield  {author} {\bibinfo {author} {\bibfnamefont {T.~F.}\ \bibnamefont
			{Watson}}, \bibinfo {author} {\bibfnamefont {S.~G.~J.}\ \bibnamefont
			{Philips}}, \bibinfo {author} {\bibfnamefont {E.}~\bibnamefont {Kawakami}},
		\bibinfo {author} {\bibfnamefont {D.~R.}\ \bibnamefont {Ward}}, \bibinfo
		{author} {\bibfnamefont {P.}~\bibnamefont {Scarlino}}, \bibinfo {author}
		{\bibfnamefont {M.}~\bibnamefont {Veldhorst}}, \bibinfo {author}
		{\bibfnamefont {D.~E.}\ \bibnamefont {Savage}}, \bibinfo {author}
		{\bibfnamefont {M.~G.}\ \bibnamefont {Lagally}}, \bibinfo {author}
		{\bibfnamefont {M.}~\bibnamefont {Friesen}}, \bibinfo {author} {\bibfnamefont
			{S.~N.}\ \bibnamefont {Coppersmith}}, \bibinfo {author} {\bibfnamefont
			{M.~A.}\ \bibnamefont {Eriksson}},\ and\ \bibinfo {author} {\bibfnamefont
			{L.~M.~K.}\ \bibnamefont {Vandersypen}},\ }\bibfield  {title} {\bibinfo
		{title} {A programmable two-qubit quantum processor in silicon},\ }\href
	{https://doi.org/10.1038/nature25766} {\bibfield  {journal} {\bibinfo
			{journal} {Nature}\ }\textbf {\bibinfo {volume} {555}},\ \bibinfo {pages}
		{633} (\bibinfo {year} {2018})}\BibitemShut {NoStop}%
	\bibitem [{\citenamefont {Dani}\ \emph {et~al.}(2022)\citenamefont {Dani},
		\citenamefont {Hussein}, \citenamefont {Bayer}, \citenamefont {Kohler},\ and\
		\citenamefont {Haug}}]{DaniCommPhys2022}%
	\BibitemOpen
	\bibfield  {author} {\bibinfo {author} {\bibfnamefont {O.}~\bibnamefont
			{Dani}}, \bibinfo {author} {\bibfnamefont {R.}~\bibnamefont {Hussein}},
		\bibinfo {author} {\bibfnamefont {J.~C.}\ \bibnamefont {Bayer}}, \bibinfo
		{author} {\bibfnamefont {S.}~\bibnamefont {Kohler}},\ and\ \bibinfo {author}
		{\bibfnamefont {R.~J.}\ \bibnamefont {Haug}},\ }\bibfield  {title} {\bibinfo
		{title} {Temperature-dependent broadening of coherent current peaks in inas
			double quantum dots},\ }\href {https://doi.org/10.1038/s42005-022-01074-z}
	{\bibfield  {journal} {\bibinfo  {journal} {Communications Physics}\ }\textbf
		{\bibinfo {volume} {5}},\ \bibinfo {pages} {292} (\bibinfo {year}
		{2022})}\BibitemShut {NoStop}%
	\bibitem [{\citenamefont {Dani}\ \emph {et~al.}(2024)\citenamefont {Dani},
		\citenamefont {Hussein}, \citenamefont {Bayer}, \citenamefont {Pierz},
		\citenamefont {Kohler},\ and\ \citenamefont {Haug}}]{DaniPRB2024}%
	\BibitemOpen
	\bibfield  {author} {\bibinfo {author} {\bibfnamefont {O.}~\bibnamefont
			{Dani}}, \bibinfo {author} {\bibfnamefont {R.}~\bibnamefont {Hussein}},
		\bibinfo {author} {\bibfnamefont {J.~C.}\ \bibnamefont {Bayer}}, \bibinfo
		{author} {\bibfnamefont {K.}~\bibnamefont {Pierz}}, \bibinfo {author}
		{\bibfnamefont {S.}~\bibnamefont {Kohler}},\ and\ \bibinfo {author}
		{\bibfnamefont {R.~J.}\ \bibnamefont {Haug}},\ }\bibfield  {title} {\bibinfo
		{title} {Direct measurement of spin-flip rates of a self-assembled inas
			double quantum dot in single-electron tunneling},\ }\href
	{https://doi.org/10.1103/PhysRevB.109.L121404} {\bibfield  {journal}
		{\bibinfo  {journal} {Phys. Rev. B}\ }\textbf {\bibinfo {volume} {109}},\
		\bibinfo {pages} {L121404} (\bibinfo {year} {2024})}\BibitemShut {NoStop}%
	\bibitem [{\citenamefont {Ubbelohde}\ \emph {et~al.}(2012)\citenamefont
		{Ubbelohde}, \citenamefont {Fricke}, \citenamefont {Flindt}, \citenamefont
		{Hohls},\ and\ \citenamefont {Haug}}]{ubbelohdeNATCOMM2012}%
	\BibitemOpen
	\bibfield  {author} {\bibinfo {author} {\bibfnamefont {N.}~\bibnamefont
			{Ubbelohde}}, \bibinfo {author} {\bibfnamefont {C.}~\bibnamefont {Fricke}},
		\bibinfo {author} {\bibfnamefont {C.}~\bibnamefont {Flindt}}, \bibinfo
		{author} {\bibfnamefont {F.}~\bibnamefont {Hohls}},\ and\ \bibinfo {author}
		{\bibfnamefont {R.~J.}\ \bibnamefont {Haug}},\ }\bibfield  {title} {\bibinfo
		{title} {Measurement of finite-frequency current statistics in a
			single-electron transistor},\ }\href {https://doi.org/10.1038/ncomms1620}
	{\bibfield  {journal} {\bibinfo  {journal} {Nat. Commun.}\ }\textbf {\bibinfo
			{volume} {3}},\ \bibinfo {pages} {612} (\bibinfo {year} {2012})}\BibitemShut
	{NoStop}%
	\bibitem [{\citenamefont {Kurzmann}\ \emph {et~al.}(2019)\citenamefont
		{Kurzmann}, \citenamefont {Stegmann}, \citenamefont {Kerski}, \citenamefont
		{Schott}, \citenamefont {Ludwig}, \citenamefont {Wieck}, \citenamefont
		{K\"onig}, \citenamefont {Lorke},\ and\ \citenamefont
		{Geller}}]{kurzmannPRL2019}%
	\BibitemOpen
	\bibfield  {author} {\bibinfo {author} {\bibfnamefont {A.}~\bibnamefont
			{Kurzmann}}, \bibinfo {author} {\bibfnamefont {P.}~\bibnamefont {Stegmann}},
		\bibinfo {author} {\bibfnamefont {J.}~\bibnamefont {Kerski}}, \bibinfo
		{author} {\bibfnamefont {R.}~\bibnamefont {Schott}}, \bibinfo {author}
		{\bibfnamefont {A.}~\bibnamefont {Ludwig}}, \bibinfo {author} {\bibfnamefont
			{A.~D.}\ \bibnamefont {Wieck}}, \bibinfo {author} {\bibfnamefont
			{J.}~\bibnamefont {K\"onig}}, \bibinfo {author} {\bibfnamefont
			{A.}~\bibnamefont {Lorke}},\ and\ \bibinfo {author} {\bibfnamefont
			{M.}~\bibnamefont {Geller}},\ }\bibfield  {title} {\bibinfo {title} {Optical
			detection of single-electron tunneling into a semiconductor quantum dot},\
	}\href {https://doi.org/10.1103/PhysRevLett.122.247403} {\bibfield  {journal}
		{\bibinfo  {journal} {Phys. Rev. Lett.}\ }\textbf {\bibinfo {volume} {122}},\
		\bibinfo {pages} {247403} (\bibinfo {year} {2019})}\BibitemShut {NoStop}%
	\bibitem [{\citenamefont {Sifft}\ \emph {et~al.}(2024)\citenamefont {Sifft},
		\citenamefont {Kurzmann}, \citenamefont {Kerski}, \citenamefont {Schott},
		\citenamefont {Ludwig}, \citenamefont {Wieck}, \citenamefont {Lorke},
		\citenamefont {Geller},\ and\ \citenamefont {H\"agele}}]{sifftPRA2024}%
	\BibitemOpen
	\bibfield  {author} {\bibinfo {author} {\bibfnamefont {M.}~\bibnamefont
			{Sifft}}, \bibinfo {author} {\bibfnamefont {A.}~\bibnamefont {Kurzmann}},
		\bibinfo {author} {\bibfnamefont {J.}~\bibnamefont {Kerski}}, \bibinfo
		{author} {\bibfnamefont {R.}~\bibnamefont {Schott}}, \bibinfo {author}
		{\bibfnamefont {A.}~\bibnamefont {Ludwig}}, \bibinfo {author} {\bibfnamefont
			{A.~D.}\ \bibnamefont {Wieck}}, \bibinfo {author} {\bibfnamefont
			{A.}~\bibnamefont {Lorke}}, \bibinfo {author} {\bibfnamefont
			{M.}~\bibnamefont {Geller}},\ and\ \bibinfo {author} {\bibfnamefont
			{D.}~\bibnamefont {H\"agele}},\ }\bibfield  {title} {\bibinfo {title}
		{Quantum polyspectra approach to the dynamics of blinking quantum emitters at
			low photon rates without binning: {Making} every photon count},\ }\href
	{https://doi.org/10.1103/PhysRevA.109.062210} {\bibfield  {journal} {\bibinfo
			{journal} {Phys. Rev. A}\ }\textbf {\bibinfo {volume} {109}},\ \bibinfo
		{pages} {062210} (\bibinfo {year} {2024})}\BibitemShut {NoStop}%
	\bibitem [{\citenamefont {Levitov}\ \emph {et~al.}(1996)\citenamefont
		{Levitov}, \citenamefont {Lee},\ and\ \citenamefont
		{Lesovik}}]{levitovJMP1996}%
	\BibitemOpen
	\bibfield  {author} {\bibinfo {author} {\bibfnamefont {L.~S.}\ \bibnamefont
			{Levitov}}, \bibinfo {author} {\bibfnamefont {H.}~\bibnamefont {Lee}},\ and\
		\bibinfo {author} {\bibfnamefont {G.~B.}\ \bibnamefont {Lesovik}},\
	}\bibfield  {title} {\bibinfo {title} {Electron counting statistics and
			coherent states of electric current},\ }\href
	{https://doi.org/10.1063/1.531672} {\bibfield  {journal} {\bibinfo  {journal}
			{J. Math. Phys.}\ }\textbf {\bibinfo {volume} {37}},\ \bibinfo {pages} {4845}
		(\bibinfo {year} {1996})}\BibitemShut {NoStop}%
	\bibitem [{\citenamefont {Bagrets}\ and\ \citenamefont
		{Nazarov}(2003)}]{bagretsPRB2003}%
	\BibitemOpen
	\bibfield  {author} {\bibinfo {author} {\bibfnamefont {D.~A.}\ \bibnamefont
			{Bagrets}}\ and\ \bibinfo {author} {\bibfnamefont {Y.~V.}\ \bibnamefont
			{Nazarov}},\ }\bibfield  {title} {\bibinfo {title} {Full counting statistics
			of charge transfer in {Coulomb} blockade systems},\ }\href@noop {} {\bibfield
		{journal} {\bibinfo  {journal} {Phys. Rev. B}\ }\textbf {\bibinfo {volume}
			{67}},\ \bibinfo {pages} {085316} (\bibinfo {year} {2003})}\BibitemShut
	{NoStop}%
	\bibitem [{\citenamefont {Flindt}\ \emph {et~al.}(2009)\citenamefont {Flindt},
		\citenamefont {Fricke}, \citenamefont {Hohls}, \citenamefont {Novotny},
		\citenamefont {Netocny}, \citenamefont {Brandes},\ and\ \citenamefont
		{Haug}}]{flindtPNAS2009}%
	\BibitemOpen
	\bibfield  {author} {\bibinfo {author} {\bibfnamefont {C.}~\bibnamefont
			{Flindt}}, \bibinfo {author} {\bibfnamefont {C.}~\bibnamefont {Fricke}},
		\bibinfo {author} {\bibfnamefont {F.}~\bibnamefont {Hohls}}, \bibinfo
		{author} {\bibfnamefont {T.}~\bibnamefont {Novotny}}, \bibinfo {author}
		{\bibfnamefont {K.}~\bibnamefont {Netocny}}, \bibinfo {author} {\bibfnamefont
			{T.}~\bibnamefont {Brandes}},\ and\ \bibinfo {author} {\bibfnamefont {R.~J.}\
			\bibnamefont {Haug}},\ }\bibfield  {title} {\bibinfo {title} {Universal
			oscillations in counting statistics},\ }\href@noop {} {\bibfield  {journal}
		{\bibinfo  {journal} {PNAS}\ }\textbf {\bibinfo {volume} {106}},\ \bibinfo
		{pages} {10116} (\bibinfo {year} {2009})}\BibitemShut {NoStop}%
	\bibitem [{\citenamefont {Cook}(1981)}]{cookPRA1981}%
	\BibitemOpen
	\bibfield  {author} {\bibinfo {author} {\bibfnamefont {R.~J.}\ \bibnamefont
			{Cook}},\ }\bibfield  {title} {\bibinfo {title} {Photon number statistics in
			resonance fluorescence},\ }\href@noop {} {\bibfield  {journal} {\bibinfo
			{journal} {Phys. Rev. A}\ }\textbf {\bibinfo {volume} {23}},\ \bibinfo
		{pages} {1243} (\bibinfo {year} {1981})}\BibitemShut {NoStop}%
	\bibitem [{\citenamefont {Kambly}\ \emph {et~al.}(2011)\citenamefont {Kambly},
		\citenamefont {Flindt},\ and\ \citenamefont {B\"uttiker}}]{kamblyPRB2011}%
	\BibitemOpen
	\bibfield  {author} {\bibinfo {author} {\bibfnamefont {D.}~\bibnamefont
			{Kambly}}, \bibinfo {author} {\bibfnamefont {C.}~\bibnamefont {Flindt}},\
		and\ \bibinfo {author} {\bibfnamefont {M.}~\bibnamefont {B\"uttiker}},\
	}\bibfield  {title} {\bibinfo {title} {Factorial cumulants reveal
			interactions in counting statistics},\ }\href
	{https://doi.org/10.1103/PhysRevB.83.075432} {\bibfield  {journal} {\bibinfo
			{journal} {Phys. Rev. B}\ }\textbf {\bibinfo {volume} {83}},\ \bibinfo
		{pages} {075432} (\bibinfo {year} {2011})}\BibitemShut {NoStop}%
	\bibitem [{\citenamefont {Stegmann}\ \emph {et~al.}(2015)\citenamefont
		{Stegmann}, \citenamefont {Sothmann}, \citenamefont {Hucht},\ and\
		\citenamefont {K\"onig}}]{stegmannPRB2015}%
	\BibitemOpen
	\bibfield  {author} {\bibinfo {author} {\bibfnamefont {P.}~\bibnamefont
			{Stegmann}}, \bibinfo {author} {\bibfnamefont {B.}~\bibnamefont {Sothmann}},
		\bibinfo {author} {\bibfnamefont {A.}~\bibnamefont {Hucht}},\ and\ \bibinfo
		{author} {\bibfnamefont {J.}~\bibnamefont {K\"onig}},\ }\bibfield  {title}
	{\bibinfo {title} {Detection of interactions via generalized factorial
			cumulants in systems in and out of equilibrium},\ }\href
	{https://doi.org/10.1103/PhysRevB.92.155413} {\bibfield  {journal} {\bibinfo
			{journal} {Phys. Rev. B}\ }\textbf {\bibinfo {volume} {92}},\ \bibinfo
		{pages} {155413} (\bibinfo {year} {2015})}\BibitemShut {NoStop}%
	\bibitem [{\citenamefont {Brange}\ \emph {et~al.}(2021)\citenamefont {Brange},
		\citenamefont {Schmidt}, \citenamefont {Bayer}, \citenamefont {Wagner},
		\citenamefont {Flindt},\ and\ \citenamefont {Haug}}]{BrangeSciAdv2021}%
	\BibitemOpen
	\bibfield  {author} {\bibinfo {author} {\bibfnamefont {F.}~\bibnamefont
			{Brange}}, \bibinfo {author} {\bibfnamefont {A.}~\bibnamefont {Schmidt}},
		\bibinfo {author} {\bibfnamefont {J.~C.}\ \bibnamefont {Bayer}}, \bibinfo
		{author} {\bibfnamefont {T.}~\bibnamefont {Wagner}}, \bibinfo {author}
		{\bibfnamefont {C.}~\bibnamefont {Flindt}},\ and\ \bibinfo {author}
		{\bibfnamefont {R.~J.}\ \bibnamefont {Haug}},\ }\bibfield  {title} {\bibinfo
		{title} {Controlled emission time statistics of a dynamic single-electron
			transistor},\ }\href {https://doi.org/10.1126/sciadv.abe0793} {\bibfield
		{journal} {\bibinfo  {journal} {Science Advances}\ }\textbf {\bibinfo
			{volume} {7}},\ \bibinfo {pages} {eabe0793} (\bibinfo {year} {2021})},\
	\Eprint
	{https://arxiv.org/abs/https://www.science.org/doi/pdf/10.1126/sciadv.abe0793}
	{https://www.science.org/doi/pdf/10.1126/sciadv.abe0793} \BibitemShut
	{NoStop}%
	\bibitem [{\citenamefont {Kleinherbers}\ \emph {et~al.}(2023)\citenamefont
		{Kleinherbers}, \citenamefont {Mannel}, \citenamefont {Kerski}, \citenamefont
		{Geller}, \citenamefont {Lorke},\ and\ \citenamefont
		{K\"onig}}]{kleinherbersPRR2023}%
	\BibitemOpen
	\bibfield  {author} {\bibinfo {author} {\bibfnamefont {E.}~\bibnamefont
			{Kleinherbers}}, \bibinfo {author} {\bibfnamefont {H.}~\bibnamefont
			{Mannel}}, \bibinfo {author} {\bibfnamefont {J.}~\bibnamefont {Kerski}},
		\bibinfo {author} {\bibfnamefont {M.}~\bibnamefont {Geller}}, \bibinfo
		{author} {\bibfnamefont {A.}~\bibnamefont {Lorke}},\ and\ \bibinfo {author}
		{\bibfnamefont {J.}~\bibnamefont {K\"onig}},\ }\bibfield  {title} {\bibinfo
		{title} {Unraveling spin dynamics from charge fluctuations},\ }\href
	{https://doi.org/10.1103/PhysRevResearch.5.043103} {\bibfield  {journal}
		{\bibinfo  {journal} {Phys. Rev. Research}\ }\textbf {\bibinfo {volume}
			{5}},\ \bibinfo {pages} {043103} (\bibinfo {year} {2023})}\BibitemShut
	{NoStop}%
	\bibitem [{\citenamefont {Emary}\ \emph {et~al.}(2007)\citenamefont {Emary},
		\citenamefont {Marcos}, \citenamefont {Aguado},\ and\ \citenamefont
		{Brandes}}]{emaryPRB2007}%
	\BibitemOpen
	\bibfield  {author} {\bibinfo {author} {\bibfnamefont {C.}~\bibnamefont
			{Emary}}, \bibinfo {author} {\bibfnamefont {D.}~\bibnamefont {Marcos}},
		\bibinfo {author} {\bibfnamefont {R.}~\bibnamefont {Aguado}},\ and\ \bibinfo
		{author} {\bibfnamefont {T.}~\bibnamefont {Brandes}},\ }\bibfield  {title}
	{\bibinfo {title} {Frequency-dependent counting statistics in interacting
			nanoscale conductors},\ }\href@noop {} {\bibfield  {journal} {\bibinfo
			{journal} {Phys. Rev. B}\ }\textbf {\bibinfo {volume} {76}},\ \bibinfo
		{pages} {161404(R)} (\bibinfo {year} {2007})}\BibitemShut {NoStop}%
	\bibitem [{\citenamefont {K\"ung}\ \emph {et~al.}(2009)\citenamefont {K\"ung},
		\citenamefont {Pf\"affli}, \citenamefont {Gustavsson}, \citenamefont {Ihn},
		\citenamefont {Ensslin}, \citenamefont {Reinwald},\ and\ \citenamefont
		{Wegscheider}}]{kungPRB2009}%
	\BibitemOpen
	\bibfield  {author} {\bibinfo {author} {\bibfnamefont {B.}~\bibnamefont
			{K\"ung}}, \bibinfo {author} {\bibfnamefont {O.}~\bibnamefont {Pf\"affli}},
		\bibinfo {author} {\bibfnamefont {S.}~\bibnamefont {Gustavsson}}, \bibinfo
		{author} {\bibfnamefont {T.}~\bibnamefont {Ihn}}, \bibinfo {author}
		{\bibfnamefont {K.}~\bibnamefont {Ensslin}}, \bibinfo {author} {\bibfnamefont
			{M.}~\bibnamefont {Reinwald}},\ and\ \bibinfo {author} {\bibfnamefont
			{W.}~\bibnamefont {Wegscheider}},\ }\bibfield  {title} {\bibinfo {title}
		{Time-resolved charge detection with cross-correlation techniques},\
	}\href@noop {} {\bibfield  {journal} {\bibinfo  {journal} {Phys. Rev. B}\
		}\textbf {\bibinfo {volume} {79}},\ \bibinfo {pages} {035314} (\bibinfo
		{year} {2009})}\BibitemShut {NoStop}%
	\bibitem [{\citenamefont {Kerski}\ \emph {et~al.}(2023)\citenamefont {Kerski},
		\citenamefont {Mannel}, \citenamefont {Lochner}, \citenamefont
		{Kleinherbers}, \citenamefont {Kurzmann}, \citenamefont {Ludwig},
		\citenamefont {Wieck}, \citenamefont {K\"onig}, \citenamefont {Lorke},\ and\
		\citenamefont {Geller}}]{kerskiSR2023}%
	\BibitemOpen
	\bibfield  {author} {\bibinfo {author} {\bibfnamefont {J.}~\bibnamefont
			{Kerski}}, \bibinfo {author} {\bibfnamefont {H.}~\bibnamefont {Mannel}},
		\bibinfo {author} {\bibfnamefont {P.}~\bibnamefont {Lochner}}, \bibinfo
		{author} {\bibfnamefont {E.}~\bibnamefont {Kleinherbers}}, \bibinfo {author}
		{\bibfnamefont {A.}~\bibnamefont {Kurzmann}}, \bibinfo {author}
		{\bibfnamefont {A.}~\bibnamefont {Ludwig}}, \bibinfo {author} {\bibfnamefont
			{A.~D.}\ \bibnamefont {Wieck}}, \bibinfo {author} {\bibfnamefont
			{J.}~\bibnamefont {K\"onig}}, \bibinfo {author} {\bibfnamefont
			{A.}~\bibnamefont {Lorke}},\ and\ \bibinfo {author} {\bibfnamefont
			{M.}~\bibnamefont {Geller}},\ }\bibfield  {title} {\bibinfo {title}
		{Post-processing of real-time quantum event measurements for an optimal
			bandwidth},\ }\href {https://doi.org/10.1038/s41598-023-28273-0} {\bibfield
		{journal} {\bibinfo  {journal} {Sci. Rep.}\ }\textbf {\bibinfo {volume}
			{13}},\ \bibinfo {pages} {1105} (\bibinfo {year} {2023})}\BibitemShut
	{NoStop}%
	\bibitem [{\citenamefont {H\"agele}\ and\ \citenamefont
		{Schefczik}(2018)}]{hagelePRB2018}%
	\BibitemOpen
	\bibfield  {author} {\bibinfo {author} {\bibfnamefont {D.}~\bibnamefont
			{H\"agele}}\ and\ \bibinfo {author} {\bibfnamefont {F.}~\bibnamefont
			{Schefczik}},\ }\bibfield  {title} {\bibinfo {title} {Higher-order moments,
			cumulants, and spectra of continuous quantum noise measurements},\
	}\href@noop {} {\bibfield  {journal} {\bibinfo  {journal} {Phys. Rev. B}\
		}\textbf {\bibinfo {volume} {98}},\ \bibinfo {pages} {205143} (\bibinfo
		{year} {2018})}\BibitemShut {NoStop}%
	\bibitem [{\citenamefont {Sifft}\ \emph {et~al.}(2021)\citenamefont {Sifft},
		\citenamefont {Kurzmann}, \citenamefont {Kerski}, \citenamefont {Schott},
		\citenamefont {Ludwig}, \citenamefont {Wieck}, \citenamefont {Lorke},
		\citenamefont {Geller},\ and\ \citenamefont {H\"agele}}]{sifftPRR2021}%
	\BibitemOpen
	\bibfield  {author} {\bibinfo {author} {\bibfnamefont {M.}~\bibnamefont
			{Sifft}}, \bibinfo {author} {\bibfnamefont {A.}~\bibnamefont {Kurzmann}},
		\bibinfo {author} {\bibfnamefont {J.}~\bibnamefont {Kerski}}, \bibinfo
		{author} {\bibfnamefont {R.}~\bibnamefont {Schott}}, \bibinfo {author}
		{\bibfnamefont {A.}~\bibnamefont {Ludwig}}, \bibinfo {author} {\bibfnamefont
			{A.~D.}\ \bibnamefont {Wieck}}, \bibinfo {author} {\bibfnamefont
			{A.}~\bibnamefont {Lorke}}, \bibinfo {author} {\bibfnamefont
			{M.}~\bibnamefont {Geller}},\ and\ \bibinfo {author} {\bibfnamefont
			{D.}~\bibnamefont {H\"agele}},\ }\bibfield  {title} {\bibinfo {title}
		{Quantum polyspectra for modeling and evaluating quantum transport
			measurements: A unifying approach to the strong and weak measurement
			regime},\ }\href {https://doi.org/10.1103/PhysRevResearch.3.033123}
	{\bibfield  {journal} {\bibinfo  {journal} {Phys. Rev. Research}\ }\textbf
		{\bibinfo {volume} {3}},\ \bibinfo {pages} {033123} (\bibinfo {year}
		{2021})}\BibitemShut {NoStop}%
	\bibitem [{\citenamefont {Landi}\ \emph {et~al.}(2024)\citenamefont {Landi},
		\citenamefont {Kewming}, \citenamefont {Mitchison},\ and\ \citenamefont
		{Potts}}]{LandiPRX2024}%
	\BibitemOpen
	\bibfield  {author} {\bibinfo {author} {\bibfnamefont {G.~T.}\ \bibnamefont
			{Landi}}, \bibinfo {author} {\bibfnamefont {M.~J.}\ \bibnamefont {Kewming}},
		\bibinfo {author} {\bibfnamefont {M.~T.}\ \bibnamefont {Mitchison}},\ and\
		\bibinfo {author} {\bibfnamefont {P.~P.}\ \bibnamefont {Potts}},\ }\bibfield
	{title} {\bibinfo {title} {Current fluctuations in open quantum systems:
			Bridging the gap between quantum continuous measurements and full counting
			statistics},\ }\href {https://doi.org/10.1103/PRXQuantum.5.020201} {\bibfield
		{journal} {\bibinfo  {journal} {PRX Quantum}\ }\textbf {\bibinfo {volume}
			{5}},\ \bibinfo {pages} {020201} (\bibinfo {year} {2024})}\BibitemShut
	{NoStop}%
	\bibitem [{\citenamefont {Jacobs}\ and\ \citenamefont
		{Steck}(2006)}]{jacobsCP2006}%
	\BibitemOpen
	\bibfield  {author} {\bibinfo {author} {\bibfnamefont {K.}~\bibnamefont
			{Jacobs}}\ and\ \bibinfo {author} {\bibfnamefont {D.~A.}\ \bibnamefont
			{Steck}},\ }\bibfield  {title} {\bibinfo {title} {A straightforward
			introduction to continuous quantum measurement},\ }\href
	{https://doi.org/10.1080/00107510601101934} {\bibfield  {journal} {\bibinfo
			{journal} {Contemp. Phys.}\ }\textbf {\bibinfo {volume} {47}},\ \bibinfo
		{pages} {279} (\bibinfo {year} {2006})}\BibitemShut {NoStop}%
	\bibitem [{\citenamefont {Korotkov}(1999)}]{korotkovPRB1999}%
	\BibitemOpen
	\bibfield  {author} {\bibinfo {author} {\bibfnamefont {A.~N.}\ \bibnamefont
			{Korotkov}},\ }\bibfield  {title} {\bibinfo {title} {Continuous quantum
			measurement of a double dot},\ }\href@noop {} {\bibfield  {journal} {\bibinfo
			{journal} {Phys. Rev. B}\ }\textbf {\bibinfo {volume} {60}},\ \bibinfo
		{pages} {5737} (\bibinfo {year} {1999})}\BibitemShut {NoStop}%
	\bibitem [{\citenamefont {Korotkov}(2001)}]{korotkovPhysRevB2001}%
	\BibitemOpen
	\bibfield  {author} {\bibinfo {author} {\bibfnamefont {A.~N.}\ \bibnamefont
			{Korotkov}},\ }\bibfield  {title} {\bibinfo {title} {Output spectrum of a
			detector measuring quantum oscillations},\ }\href
	{https://doi.org/10.1103/PhysRevB.63.085312} {\bibfield  {journal} {\bibinfo
			{journal} {Phys. Rev. B}\ }\textbf {\bibinfo {volume} {63}},\ \bibinfo
		{pages} {085312} (\bibinfo {year} {2001})}\BibitemShut {NoStop}%
	\bibitem [{\citenamefont {Barchielli}\ \emph {et~al.}(1982)\citenamefont
		{Barchielli}, \citenamefont {Lanz},\ and\ \citenamefont
		{Prosperi}}]{barchielliNC1982}%
	\BibitemOpen
	\bibfield  {author} {\bibinfo {author} {\bibfnamefont {A.}~\bibnamefont
			{Barchielli}}, \bibinfo {author} {\bibfnamefont {L.}~\bibnamefont {Lanz}},\
		and\ \bibinfo {author} {\bibfnamefont {G.~M.}\ \bibnamefont {Prosperi}},\
	}\bibfield  {title} {\bibinfo {title} {A model for the macroscopic
			description and continual observations in quantum mechanics},\ }\href@noop {}
	{\bibfield  {journal} {\bibinfo  {journal} {Nuovo Cimento}\ }\textbf
		{\bibinfo {volume} {72B}},\ \bibinfo {pages} {79} (\bibinfo {year}
		{1982})}\BibitemShut {NoStop}%
	\bibitem [{\citenamefont {Sifft}\ and\ \citenamefont
		{H\"agele}(2023)}]{SifftPRA2023}%
	\BibitemOpen
	\bibfield  {author} {\bibinfo {author} {\bibfnamefont {M.}~\bibnamefont
			{Sifft}}\ and\ \bibinfo {author} {\bibfnamefont {D.}~\bibnamefont
			{H\"agele}},\ }\bibfield  {title} {\bibinfo {title} {Random-time quantum
			measurements},\ }\href {https://doi.org/10.1103/PhysRevA.107.052203}
	{\bibfield  {journal} {\bibinfo  {journal} {Phys. Rev. A}\ }\textbf {\bibinfo
			{volume} {107}},\ \bibinfo {pages} {052203} (\bibinfo {year}
		{2023})}\BibitemShut {NoStop}%
	\bibitem [{\citenamefont {Maisi}\ \emph {et~al.}(2016)\citenamefont {Maisi},
		\citenamefont {Hofmann}, \citenamefont {R\"o\"osli}, \citenamefont {Basset},
		\citenamefont {Reichl}, \citenamefont {Wegscheider}, \citenamefont {Ihn},\
		and\ \citenamefont {Ensslin}}]{MaisiPRL2016}%
	\BibitemOpen
	\bibfield  {author} {\bibinfo {author} {\bibfnamefont {V.~F.}\ \bibnamefont
			{Maisi}}, \bibinfo {author} {\bibfnamefont {A.}~\bibnamefont {Hofmann}},
		\bibinfo {author} {\bibfnamefont {M.}~\bibnamefont {R\"o\"osli}}, \bibinfo
		{author} {\bibfnamefont {J.}~\bibnamefont {Basset}}, \bibinfo {author}
		{\bibfnamefont {C.}~\bibnamefont {Reichl}}, \bibinfo {author} {\bibfnamefont
			{W.}~\bibnamefont {Wegscheider}}, \bibinfo {author} {\bibfnamefont
			{T.}~\bibnamefont {Ihn}},\ and\ \bibinfo {author} {\bibfnamefont
			{K.}~\bibnamefont {Ensslin}},\ }\bibfield  {title} {\bibinfo {title}
		{Spin-orbit coupling at the level of a single electron},\ }\href
	{https://doi.org/10.1103/PhysRevLett.116.136803} {\bibfield  {journal}
		{\bibinfo  {journal} {Phys. Rev. Lett.}\ }\textbf {\bibinfo {volume} {116}},\
		\bibinfo {pages} {136803} (\bibinfo {year} {2016})}\BibitemShut {NoStop}%
	\bibitem [{\citenamefont {Bayer}\ \emph {et~al.}(2019)\citenamefont {Bayer},
		\citenamefont {Wagner}, \citenamefont {Rugeramigabo},\ and\ \citenamefont
		{Haug}}]{BayerAnnPhys2019}%
	\BibitemOpen
	\bibfield  {author} {\bibinfo {author} {\bibfnamefont {J.~C.}\ \bibnamefont
			{Bayer}}, \bibinfo {author} {\bibfnamefont {T.}~\bibnamefont {Wagner}},
		\bibinfo {author} {\bibfnamefont {E.~P.}\ \bibnamefont {Rugeramigabo}},\ and\
		\bibinfo {author} {\bibfnamefont {R.~J.}\ \bibnamefont {Haug}},\ }\bibfield
	{title} {\bibinfo {title} {Charge reconfiguration in isolated quantum dot
			arrays},\ }\href {https://doi.org/https://doi.org/10.1002/andp.201800393}
	{\bibfield  {journal} {\bibinfo  {journal} {Annalen der Physik}\ }\textbf
		{\bibinfo {volume} {531}},\ \bibinfo {pages} {1800393} (\bibinfo {year}
		{2019})},\ \Eprint
	{https://arxiv.org/abs/https://onlinelibrary.wiley.com/doi/pdf/10.1002/andp.201800393}
	{https://onlinelibrary.wiley.com/doi/pdf/10.1002/andp.201800393} \BibitemShut
	{NoStop}%
	\bibitem [{\citenamefont {Brillinger}(1965)}]{brillingerAMS1965}%
	\BibitemOpen
	\bibfield  {author} {\bibinfo {author} {\bibfnamefont {D.~R.}\ \bibnamefont
			{Brillinger}},\ }\bibfield  {title} {\bibinfo {title} {An introduction to
			polyspectra},\ }\href@noop {} {\bibfield  {journal} {\bibinfo  {journal}
			{Ann. Math. Statist.}\ }\textbf {\bibinfo {volume} {36}},\ \bibinfo {pages}
		{1351} (\bibinfo {year} {1965})}\BibitemShut {NoStop}%
	\bibitem [{\citenamefont {Gardiner}(2009)}]{gardinerBOOK2009}%
	\BibitemOpen
	\bibfield  {author} {\bibinfo {author} {\bibfnamefont {C.}~\bibnamefont
			{Gardiner}},\ }\href@noop {} {\emph {\bibinfo {title} {Stochastic
				Methods}}},\ \bibinfo {edition} {4th}\ ed.\ (\bibinfo  {publisher}
	{Springer},\ \bibinfo {address} {Berlin Heidelberg},\ \bibinfo {year}
	{2009})\BibitemShut {NoStop}%
	\bibitem [{\citenamefont {Sifft}(2022)}]{sifftSIGNALSNAP2022}%
	\BibitemOpen
	\bibfield  {author} {\bibinfo {author} {\bibfnamefont {M.}~\bibnamefont
			{Sifft}},\ }\href@noop {} {\bibinfo {title} {{SignalSnap Toolbox}}},\
	\bibinfo {howpublished} {\url{https://github.com/MarkusSifft/SignalSnap}}
	(\bibinfo {year} {2022})\BibitemShut {NoStop}%
	\bibitem [{\citenamefont {Schefczik}\ and\ \citenamefont
		{H\"agele}(2019)}]{schefczikARXIV2019}%
	\BibitemOpen
	\bibfield  {author} {\bibinfo {author} {\bibfnamefont {F.}~\bibnamefont
			{Schefczik}}\ and\ \bibinfo {author} {\bibfnamefont {D.}~\bibnamefont
			{H\"agele}},\ }\bibfield  {title} {\bibinfo {title} {{Ready-to-Use Unbiased
				Estimators for Multivariate Cumulants Including One That Outperforms
				$\bar{x^3}$}},\ }\href@noop {} {\  (\bibinfo {year} {2019})},\ \Eprint
	{https://arxiv.org/abs/1904.12154} {arXiv:1904.12154 [math.st]} \BibitemShut
	{NoStop}%
	\bibitem [{\citenamefont {Tilloy}(2018)}]{tilloyPRA2018}%
	\BibitemOpen
	\bibfield  {author} {\bibinfo {author} {\bibfnamefont {A.}~\bibnamefont
			{Tilloy}},\ }\bibfield  {title} {\bibinfo {title} {Exact signal correlators
			in continuous quantum measurements},\ }\href@noop {} {\bibfield  {journal}
		{\bibinfo  {journal} {Phys. Rev. A}\ }\textbf {\bibinfo {volume} {98}},\
		\bibinfo {pages} {010104(R)} (\bibinfo {year} {2018})}\BibitemShut {NoStop}%
	\bibitem [{foo()}]{footnote1}%
	\BibitemOpen
	\href@noop {} {\bibinfo {title} {{Please note, that $k$ is not required on
				the right-hand side.}}}\BibitemShut {Stop}%
	\bibitem [{\citenamefont {Sifft}(2023{\natexlab{a}})}]{sifftQUANTUMCATCH2023}%
	\BibitemOpen
	\bibfield  {author} {\bibinfo {author} {\bibfnamefont {M.}~\bibnamefont
			{Sifft}},\ }\href@noop {} {\bibinfo {title} {{QuantumCatch Toolbox}}},\
	\bibinfo {howpublished} {\url{https://github.com/MarkusSifft/QuantumCatch}}
	(\bibinfo {year} {2023}{\natexlab{a}})\BibitemShut {NoStop}%
	\bibitem [{\citenamefont
		{Sifft}(2023{\natexlab{b}})}]{sifftMarkovAnalyzer2023}%
	\BibitemOpen
	\bibfield  {author} {\bibinfo {author} {\bibfnamefont {M.}~\bibnamefont
			{Sifft}},\ }\href@noop {} {\bibinfo {title} {{SignalSnap Toolbox}}},\
	\bibinfo {howpublished} {\url{https://github.com/MarkusSifft/MarkovAnalyzer}}
	(\bibinfo {year} {2023}{\natexlab{b}})\BibitemShut {NoStop}%
	\bibitem [{\citenamefont {Yalamanchili}\ \emph {et~al.}(2015)\citenamefont
		{Yalamanchili}, \citenamefont {Arshad}, \citenamefont {Mohammed},
		\citenamefont {Garigipati}, \citenamefont {Entschev}, \citenamefont
		{Kloppenborg}, \citenamefont {Malcolm},\ and\ \citenamefont
		{Melonakos}}]{Yalamanchili2015}%
	\BibitemOpen
	\bibfield  {author} {\bibinfo {author} {\bibfnamefont {P.}~\bibnamefont
			{Yalamanchili}}, \bibinfo {author} {\bibfnamefont {U.}~\bibnamefont
			{Arshad}}, \bibinfo {author} {\bibfnamefont {Z.}~\bibnamefont {Mohammed}},
		\bibinfo {author} {\bibfnamefont {P.}~\bibnamefont {Garigipati}}, \bibinfo
		{author} {\bibfnamefont {P.}~\bibnamefont {Entschev}}, \bibinfo {author}
		{\bibfnamefont {B.}~\bibnamefont {Kloppenborg}}, \bibinfo {author}
		{\bibfnamefont {J.}~\bibnamefont {Malcolm}},\ and\ \bibinfo {author}
		{\bibfnamefont {J.}~\bibnamefont {Melonakos}},\ }\href
	{https://github.com/arrayfire/arrayfire} {\bibinfo {title} {{ArrayFire - A
				high performance software library for parallel computing with an easy-to-use
				API}}} (\bibinfo {year} {2015})\BibitemShut {NoStop}%
	\bibitem [{\citenamefont {Akaike}(1974)}]{AkaikeIEEE1974}%
	\BibitemOpen
	\bibfield  {author} {\bibinfo {author} {\bibfnamefont {H.}~\bibnamefont
			{Akaike}},\ }\bibfield  {title} {\bibinfo {title} {A new look at the
			statistical model identification},\ }\href
	{https://doi.org/10.1109/TAC.1974.1100705} {\bibfield  {journal} {\bibinfo
			{journal} {IEEE Transactions on Automatic Control}\ }\textbf {\bibinfo
			{volume} {19}},\ \bibinfo {pages} {716} (\bibinfo {year} {1974})}\BibitemShut
	{NoStop}%
	\bibitem [{\citenamefont {Bayer}\ \emph {et~al.}(2024)\citenamefont {Bayer},
		\citenamefont {Brange}, \citenamefont {Schmidt}, \citenamefont {Wagner},
		\citenamefont {Rugeramigabo}, \citenamefont {Flindt},\ and\ \citenamefont
		{Haug}}]{BayerArxiv2024}%
	\BibitemOpen
	\bibfield  {author} {\bibinfo {author} {\bibfnamefont {J.~C.}\ \bibnamefont
			{Bayer}}, \bibinfo {author} {\bibfnamefont {F.}~\bibnamefont {Brange}},
		\bibinfo {author} {\bibfnamefont {A.}~\bibnamefont {Schmidt}}, \bibinfo
		{author} {\bibfnamefont {T.}~\bibnamefont {Wagner}}, \bibinfo {author}
		{\bibfnamefont {E.~P.}\ \bibnamefont {Rugeramigabo}}, \bibinfo {author}
		{\bibfnamefont {C.}~\bibnamefont {Flindt}},\ and\ \bibinfo {author}
		{\bibfnamefont {R.~J.}\ \bibnamefont {Haug}},\ }\bibfield  {title} {\bibinfo
		{title} {Real-time detection and control of correlated charge tunneling in a
			quantum dot},\ }\href {https://arxiv.org/abs/2405.16724} {\  (\bibinfo {year}
		{2024})},\ \Eprint {https://arxiv.org/abs/2405.16724} {arXiv:2405.16724}
	\BibitemShut {NoStop}%
	\bibitem [{\citenamefont {H\"agele}\ \emph {et~al.}(2020)\citenamefont
		{H\"agele}, \citenamefont {Sifft},\ and\ \citenamefont
		{Schefczik}}]{hagelePRB2020E}%
	\BibitemOpen
	\bibfield  {author} {\bibinfo {author} {\bibfnamefont {D.}~\bibnamefont
			{H\"agele}}, \bibinfo {author} {\bibfnamefont {M.}~\bibnamefont {Sifft}},\
		and\ \bibinfo {author} {\bibfnamefont {F.}~\bibnamefont {Schefczik}},\
	}\bibfield  {title} {\bibinfo {title} {Erratum: {Higher-order} moments,
			cumulants, and spectra of continuous quantum noise measurements { [Phys. Rev.
				B 98, 205143 (2018)]}},\ }\href@noop {} {\bibfield  {journal} {\bibinfo
			{journal} {Phys. Rev. B}\ }\textbf {\bibinfo {volume} {102}},\ \bibinfo
		{pages} {119901(E)} (\bibinfo {year} {2020})}\BibitemShut {NoStop}%
	\bibitem [{\citenamefont {Johansson}\ \emph {et~al.}(2013)\citenamefont
		{Johansson}, \citenamefont {Nation},\ and\ \citenamefont
		{Nori}}]{JOHANSSON20131234}%
	\BibitemOpen
	\bibfield  {author} {\bibinfo {author} {\bibfnamefont {J.}~\bibnamefont
			{Johansson}}, \bibinfo {author} {\bibfnamefont {P.}~\bibnamefont {Nation}},\
		and\ \bibinfo {author} {\bibfnamefont {F.}~\bibnamefont {Nori}},\ }\bibfield
	{title} {\bibinfo {title} {Qutip 2: A {Python} framework for the dynamics of
			open quantum systems},\ }\href
	{https://doi.org/https://doi.org/10.1016/j.cpc.2012.11.019} {\bibfield
		{journal} {\bibinfo  {journal} {Comput. Phys. Commun.}\ }\textbf {\bibinfo
			{volume} {184}},\ \bibinfo {pages} {1234 } (\bibinfo {year}
		{2013})}\BibitemShut {NoStop}%
	\bibitem [{\citenamefont {Brandes}(2008)}]{Brandes2008}%
	\BibitemOpen
	\bibfield  {author} {\bibinfo {author} {\bibfnamefont {T.}~\bibnamefont
			{Brandes}},\ }\bibfield  {title} {\bibinfo {title} {Waiting times and noise
			in single particle transport},\ }\href
	{https://doi.org/https://doi.org/10.1002/andp.20085200707} {\bibfield
		{journal} {\bibinfo  {journal} {Annalen der Physik}\ }\textbf {\bibinfo
			{volume} {520}},\ \bibinfo {pages} {477} (\bibinfo {year} {2008})},\ \Eprint
	{https://arxiv.org/abs/https://onlinelibrary.wiley.com/doi/pdf/10.1002/andp.20085200707}
	{https://onlinelibrary.wiley.com/doi/pdf/10.1002/andp.20085200707}
	\BibitemShut {NoStop}%
\end{thebibliography}
\end{document}